\documentclass[english,aps,pra,superscriptaddress,address,showacknowledgments]{revtex4}
\pdfoutput=1
\usepackage[latin9]{inputenc}
\usepackage{color}
\usepackage{babel}
\usepackage{amsmath}
\usepackage{amssymb}
\usepackage{graphicx}
\usepackage{esint}
\usepackage[unicode=true,
 bookmarks=true,bookmarksnumbered=false,bookmarksopen=false,
 breaklinks=false,pdfborder={0 0 1},backref=false,colorlinks=false]
 {hyperref}
\hypersetup{pdftitle={Noise-Induced Transitions in Optomechanical Synchronization},
 pdfauthor={Talitha Weiss, Andreas Kronwald, and Florian Marquardt}}

\makeatletter
\@ifundefined{textcolor}{}
{%
 \definecolor{BLACK}{gray}{0}
 \definecolor{WHITE}{gray}{1}
 \definecolor{RED}{rgb}{1,0,0}
 \definecolor{GREEN}{rgb}{0,1,0}
 \definecolor{BLUE}{rgb}{0,0,1}
 \definecolor{CYAN}{cmyk}{1,0,0,0}
 \definecolor{MAGENTA}{cmyk}{0,1,0,0}
 \definecolor{YELLOW}{cmyk}{0,0,1,0}
}

\usepackage{babel}
\usepackage{bbold}

\hypersetup{
   colorlinks=true,       % false: boxed links; true: colored links
    linkcolor=blue,          % color of internal links (change box color with linkbordercolor)
    citecolor=blue,        % color of links to bibliography
    filecolor=magenta,      % color of file links
    urlcolor=blue           % color of external links
}

\makeatother

\begin{document}

\title{Noise-Induced Transitions in Optomechanical Synchronization}

\author{Talitha Weiss}

\affiliation{Friedrich-Alexander University Erlangen-Nürnberg (FAU), Department
of Physics, Staudtstr. 7, 91058 Erlangen, Germany}

\author{Andreas Kronwald}

\affiliation{Friedrich-Alexander University Erlangen-Nürnberg (FAU), Department
of Physics, Staudtstr. 7, 91058 Erlangen, Germany}

\author{Florian Marquardt}

\affiliation{Friedrich-Alexander University Erlangen-Nürnberg (FAU), Department
of Physics, Staudtstr. 7, 91058 Erlangen, Germany}

\affiliation{Max Planck Institute for the Science of Light, Günther-Scharowsky-Straße
1/Bau 24, 91058 Erlangen, Germany}
\begin{abstract}
We study how quantum and thermal noise affects synchronization of
two optomechanical limit-cycle oscillators. Classically, in the absence
of noise, optomechanical systems tend to synchronize either in-phase
or anti-phase. Taking into account the fundamental quantum noise,
we find a regime where fluctuations drive transitions between these
classical synchronization states. We investigate how this ``mixed''
synchronization regime emerges from the noiseless system by studying
the classical-to-quantum crossover and we show how the time scales
of the transitions vary with the effective noise strength. In addition,
we compare the effects of thermal noise to the effects of quantum
noise.
\end{abstract}
\maketitle

\section{Introduction}

The field of cavity optomechanics deals with systems where light in
an optical cavity couples to mechanical motion \cite{2014_Review}.
Recent experimental progress allows to move on from the investigation
of a single optomechanical system to several coupled optomechanical
systems. There are already first experiments that involve a few mechanical
and optical modes \cite{2012_Hill_CoherentWavelengthConv,2012_Dong_OMDarkMode,2010_Grudinin_PhononLaserAction,2012_Bahl_SpontaneousBrillouinCooling},
exploiting them for wavelength conversion, phonon lasing and efficient
cooling. Such few-mode optomechanical setups have been the subject
of an increasing number of theoretical proposals, on topics such as
efficient state transfer \cite{2012_Wang_QuantumStateTransfer}, two-mode
squeezing \cite{2014_Woolley_TwoModeSqueezing}, back-action evading
measurements \cite{2013_Woolley_BAEmsmt}, entanglement \cite{2007_Paternostro_EntanglementWithLight,2014_Deng_CV_entanglement,2015_Wang_OutputEntanglement_3ModeOM}
or Landau-Zener dynamics \cite{2010_Heinrich_PhotonShuttle,2013_Wu_LandauZenerPhononLasing}. 

\begin{figure}
\centering{}\includegraphics[scale=0.4]{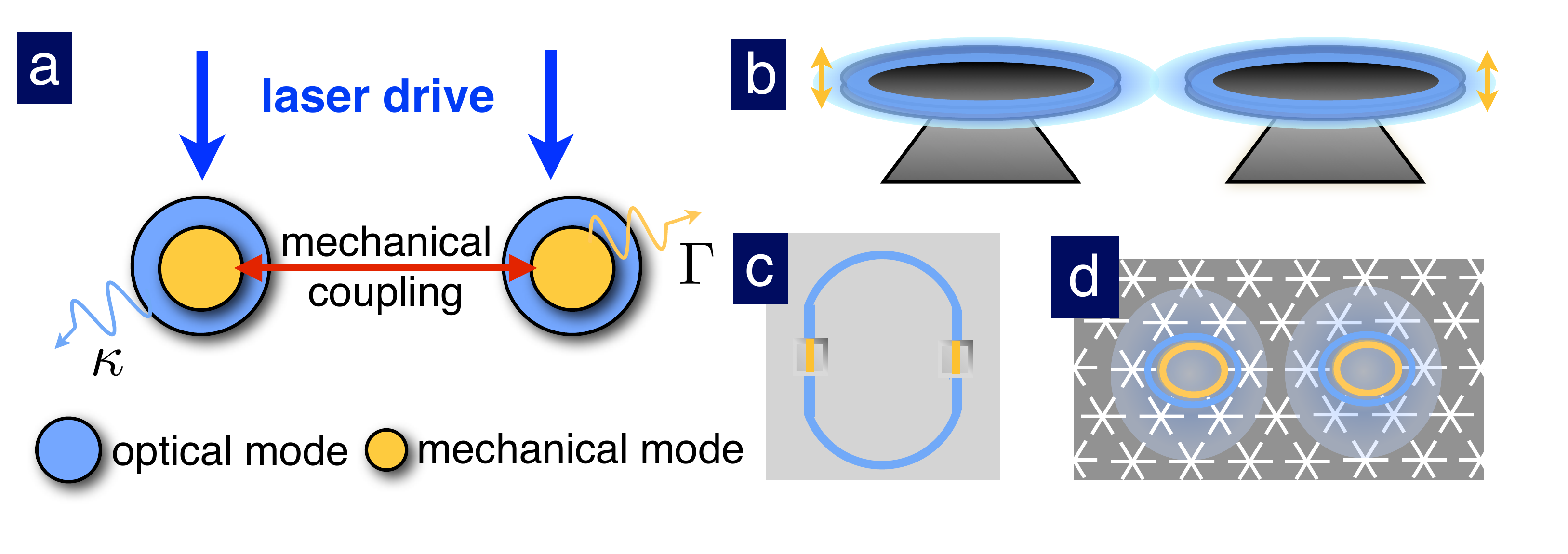}\protect\caption{Synchronization of optomechanical oscillators. Schematics of the setup
we study (a) and possible experimental implementations (b)-(d). (b)
Micro-disk oscillators that support optical whispering gallery modes
which couple evanescently to each other \cite{2012_ZhangLipson_SynchronizationPRL,2015_Lipson_ArraySync},
(c) optical racetrack resonator coupled to two nanomechanical oscillators
\cite{2013_BagheriTang_Synchronization}, (d) optomechanical crystal
structure with two optomechanical cells in the vicinity of each other,
allowing for optical and mechanical coupling \cite{2014_Safavi-Naeini_OmCrystal}.\label{fig_Setup}}
\end{figure}

Larger arrays may be implemented using a variety of settings, such
as coupled disks \cite{2012_ZhangLipson_SynchronizationPRL} (Fig.~\ref{fig_Setup}(b))
or optomechanical crystal structures \cite{2006_Maldovan_OMcrystals,2009_Eichenfield_OptomechanicalCrystals,2010_Safavi-Naeini_Slotted2D,2011_Sagnes2DCrystaldefect,2011_Laude_phoxonicCrystal,2014_Safavi-Naeini_OmCrystal}
(Fig.~\ref{fig_Setup}(d)). Optomechanical arrays have also attracted
attention from a theoretical point of view. They have been studied
in the context of slowing light \cite{2011_Chang_SlowingAndStoppingLight_NJP},
Dirac physics \cite{2015_Schmidt_OMDiracPhysics}, reservoir engineering
\cite{2012_TomadinZoller_ReservoirEngineeringOptomechanicalArray},
artificial magnetic fields for photons \cite{2015_Schmidt_MagneticFieldForPhotons},
heat transport \cite{2015_Xuereb_HeatTransport_Array}, and topological
phases of sound and light \cite{2014_Peano_TopologicalPhases}. Furthermore,
multi-membrane systems \cite{2008_Bhattacharya_OM_multiMembrane,2012_Xuereb_MembraneOptomechanicalArray,2013_Xuereb_EnhancedOMCoupling_Array,2014_Xuereb_OMArray_LongRangePhononDynamics}
were studied theoretically, considering for instance long-range interactions
and dynamics.

Most notably, optomechanical arrays provide a platform to study synchronization
of mechanical oscillators. This was initially pointed out in Ref.~\cite{2011_Heinrich_CollectiveDynamics}.
Synchronization is a well known phenomenon in many different branches
of science \cite{2001_Kurths_Synchronization} and typically arises whenever there are stable limit-cycle oscillations.
However, we note that synchronization-like phenomena have been recently studied also in the context of linear oscillators dissipating into a common bath \cite{2012_Giorgi_QCorrAndSync}. 
Optomechanical systems exhibit a Hopf bifurcation and can be optically driven into
mechanical limit-cycle oscillations \cite{2004_KarraiConstanze_IEEE,2005_KippenbergVahala_TheoryPRL,2006_FM_DynamicalMultistability,2008_Metzger}.
These self-oscillations have also been analyzed theoretically in the
quantum regime \cite{2008_LudwigMarquardt_OptomechInstab,2012_QianFM_NonclassicalStates,2014_Loerch_QLimitCycles}.
The theoretical description of the synchronization dynamics in optomechanical
arrays has initially focussed on the classical regime \cite{2011_Heinrich_CollectiveDynamics,2012_Holmes_Synchronization}.
More recent insights into this regime include the pattern formation
of the mechanical phase field in larger optomechanical arrays \cite{2014_Lauter_PhasePatterns}.
In the quantum regime, it was found \cite{2013_Ludwig} that quantum
noise can drive a sharp nonequilibrium transition towards an unsynchronized
state in an extended array, even for optomechanical systems with identical
frequencies. Further general insights into quantum synchronization
were gained in a model system of one van-der-Pol oscillator coupled
to an external drive \cite{2014_Walter_QuantumSync_1Vdp} or two coupled
van-der-Pol oscillators \cite{2014_Walter_SynchronizationVdPosc},
which can serve as a rough approximation to an actual optomechanical
system. A number of more recent works have explored quantum synchronization
on the more conceptual level \cite{2014_HermosoDeMendoza_SyncSemiclassicalKuramoto},
as well as in various other physical systems, such as e.g.~trapped
atoms and ions \cite{2013_Lee_QuantumSync_VdP_Ions,2014_Lee_EntanglementTongue_QSync,2014_Hush_SpinCorr_QSync,2014_Minghui_QuantumSync_Atoms},
qubits \cite{2009_QSync_Qubits,2012_Savelev_QSyncACdrives,2013_Giorgi_QSync_Spins},
and superconducting devices \cite{2013_Hriscu_QSync_SuperconductingDevice}. The relation of synchronization and correlations in the quantum-to-classical transition was studied in a system of coupled cavities containing a non-linearity \cite{2013_Lee_CavSync_QCL}.
Notably, it is still challenging to define a good measure for quantum
synchronization \cite{2013_Mari_QuantumSync,2015_Ameri_MutualInformation_QuantumSync,2014_Hush_SpinCorr_QSync}. 

Only recently synchronization of two nanomechanical oscillators in
the classical regime was demonstrated experimentally. This has been
achieved using optomechanical systems with coupled micro-disks \cite{2012_ZhangLipson_SynchronizationPRL}
(Fig.~\ref{fig_Setup}(b)), as well as in an experiment involving
an optical racetrack cavity coupled to two mechanical oscillators
\cite{2013_BagheriTang_Synchronization} (Fig.~\ref{fig_Setup}(c)),
and also in a setup using nanoelectromechanical systems \cite{2014_Matheny_PhaseSyncExp}.
In a recent first step towards larger arrays, up to seven optically
coupled micro-disks were used to demonstrate the expected phase noise
reduction due to synchronization \cite{2015_Lipson_ArraySync}. This $1/N$ phase noise reduction with the number of coupled systems $N$ is considered to be one of the main prospects of synchronized nanomechanical arrays. Indeed, recognized from the very beginning with the synchronization of pendulum clocks \cite{1673_Huygens}, synchronization has the potential to improve time-keeping and frequency stability. Examples where different types of synchronization have been applied or suggested for application are for frequency stabilization of high power lasers by coupling to a more stable, low power laser \cite{1972_Buczek_AppliedSync_Lasers} and for secure communication in connection with chaos \cite{1993_Cuomo_SyncAppl_Comm}. For a more complete overview and also the many applications to biology see e.g.~\cite{2001_Kurths_Synchronization,2003_Rosenblum_SyncRev,2014_Doerfler_SyncRev}.

In this work we study the effects of quantum and thermal noise on the synchronization of two optomechanical systems.  We focus on a bistable synchronization regime that either exists already in the absence of noise, or is induced by it. Bistabilities in quantum systems have been investigated before \cite{1988_Dykman_BistabQuantum,2006_Peano_Bistab_QuantumDuffing,2011_Ghobadi_Qom_bistab} and noise-induced bistabilities are known in several other systems in biology and chemistry \cite{ 2004_NoiseIndBistab_bio, 2014_Biancalani_NoiseIndBistab_bio,2015_BistabNoise_exact}, as well as in physics \cite{1989_Fichthorn_Bistab_MC, 1996_Kim_Multistab_OscArr, 2001_Residori_Bistab_SurfWaves}. We find and discuss noise-induced bistable behaviour now in the context of optomechanical (quantum) synchronization. 

This manuscript is organized as follows: We begin with a brief
review of classical synchronization in the absence of noise, Sec.~\ref{sec_CLsync}.
Then, we introduce our model in Sec.~\ref{sec_Model}, explain our methods in Sec.~\ref{sec_Methods}, and state our
main results in Sec.~\ref{sec_Results}. We note that both quantum and thermal noise lead to similar effects, but start out with the investigation of quantum noise effects.
Therefore, in Sec.~\ref{sec_FullQ},
we simulate the full quantum behaviour of the system, in contrast
to the previous investigation presented in Ref.~\cite{2013_Ludwig}. We find a regime of ``mixed'' synchronization
with two stable synchronization states, and we explore its classical-to-quantum
crossover in Sec.~\ref{sec_QtoCL}. In Sec.~\ref{sec_SyncRegimes},
we give an overview of the different synchronization regimes. Finally,
in Sec.~\ref{sec_ThNoise}, we discuss the effects of thermal noise
as compared to quantum noise. This is important to gauge the potential
of observing the quantum noise effects discussed here in future experiments.

\section{Brief review: Classical Synchronization of optomechanical Oscillators\label{sec_CLsync}}

In this section we briefly review the concepts of classical synchronization
of optomechanical systems in the absence of noise. This will set the
stage for the discussion of our results on synchronization in the
presence of thermal and quantum noise. 

A widely studied model for synchronization is the Kuramoto model \cite{1975_Kuramoto_original,2005_Acebron_KuramotoReview},
which describes a set of coupled phase oscillators. Each phase oscillator
has a phase $\phi_{i}$ and an intrinsic frequency $\omega_{i}$,
and couples to the other oscillators via the phase difference. For
two oscillators, the two corresponding phase equations collapse into
one equation for the relative phase $\delta\phi=\phi_{2}-\phi_{1}$,

\begin{equation}
\delta\dot{\phi}=(\omega_{2}-\omega_{1})-k\sin(\delta\phi).\label{eq_Kuramoto}
\end{equation}
The two oscillators are synchronized if their respective phase velocities
become equal, $\dot{\phi}_{1}=\dot{\phi}_{2}$, i.e.~$\delta\phi=\text{\text{const}}$.
From Eq.~(\ref{eq_Kuramoto}) one finds the synchronization threshold:
The two oscillators are synchronized if the coupling $k$ exceeds
the natural frequency difference, $|k|>|\omega_{2}-\omega_{1}|$.
Following this condition, oscillators with identical intrinsic frequencies
$\omega_{i}$ are always synchronized. There is only a single stable
value of $\delta\phi$ in the synchronized regime. For $k\rightarrow+\infty$,
this value approaches zero, $\delta\phi\rightarrow0$, whereas $\delta\phi\rightarrow\pi$
for $k\rightarrow-\infty$. 

Although synchronization appears in a large range of systems which
are very different in terms of microscopic parameters, their behaviour
can often be captured by effective phase equations of the Kuramoto-type
\cite{2001_Kurths_Synchronization,2005_Acebron_KuramotoReview}. In
the context of optomechanics, the mechanical motion of a single optomechanical
system near the Hopf bifurcation can be described with a phase and
an amplitude equation \cite{2006_FM_DynamicalMultistability}. Starting
from these equations, an effective Kuramoto-type model for coupled
optomechanical systems has been derived \cite{2011_Heinrich_CollectiveDynamics}.
This model describes arrays of arbitrary many optomechanical cells
with arbitrary intrinsic frequencies. For two oscillators, the phase
equations of this model reduce, again, to a single equation for the
phase difference 

\begin{equation}
\delta\dot{\phi}=(\omega_{2}-\omega_{1})-2S_{1}\sin(\delta\phi)-4S_{2}\sin(2\delta\phi).\label{eq_HopfKuramoto}
\end{equation}
Here, $S_{1}$ and $S_{2}$ are effective parameters depending on
the microscopic parameters of the underlying optomechanical systems
\cite{2011_Heinrich_CollectiveDynamics,2013_Ludwig}. Note that the
$S_{1}$-term was added to the model only later \cite{2013_Ludwig},
and accounts for the change of the intrinsic frequencies $\omega_{i}$
with the mechanical oscillation amplitude. Recently, the resulting
full Hopf-Kuramoto model has been used to study pattern formation
in 2D arrays of optomechanical systems \cite{2014_Lauter_PhasePatterns}.
In contrast to the original Kuramoto model Eq.~(\ref{eq_Kuramoto}),
the optomechanical Hopf-Kuramoto model Eq.~(\ref{eq_HopfKuramoto})
includes a term that involves $\sin(2\delta\phi)$. Rewriting Eq.~(\ref{eq_HopfKuramoto})
in terms of an effective potential, $\delta\dot{\phi}=-U'(\delta\phi)$
\cite{2011_Heinrich_CollectiveDynamics}, this term corresponds to
the appearance of a second minimum close to $\delta\phi=\pi$ which
can co-exist with the minimum close to $\delta\phi=0$. This allows
optomechanical systems to synchronize not only in-phase ($0$-synchronization),
$\delta\phi\rightarrow0$, but also anti-phase ($\pi$-synchronization),
$\delta\phi\rightarrow\pi$. For the effective potential there are
three different possibilities, depending on the parameters $S_{1}$
and $S_{2}$: (i) It has a single minimum close to $\delta\phi=0$
which leads to $0$-synchronization only, (ii) it has a single minimum
close to $\delta\phi=\pi$ which leads to $\pi$-synchronization only,
or (iii) both minima appear simultaneously in the effective potential,
such that the initial conditions determine whether $0$- or $\pi$-synchronization
occurs. These three regimes are schematically shown in Fig.~\ref{fig_Threshold}(b) and (c)
and discussed in Sec.~\ref{sec_Model}.

In general, for optomechanical arrays an effective potential for the
phases $\phi_{i}$ does not exist \cite{2014_Lauter_PhasePatterns}.
However, in the case of two oscillators only, the system can be described
with one degree of freedom, i.e.~the relative phase $\delta\phi$,
and an effective potential for $\delta\phi$ can always be constructed.
Below, we make use of the existence of this effective potential to
give an intuitive understanding of synchronization even in the presence
of noise.

\section{Model\label{sec_Model}}

\begin{figure}
\begin{centering}
\includegraphics[scale=0.38]{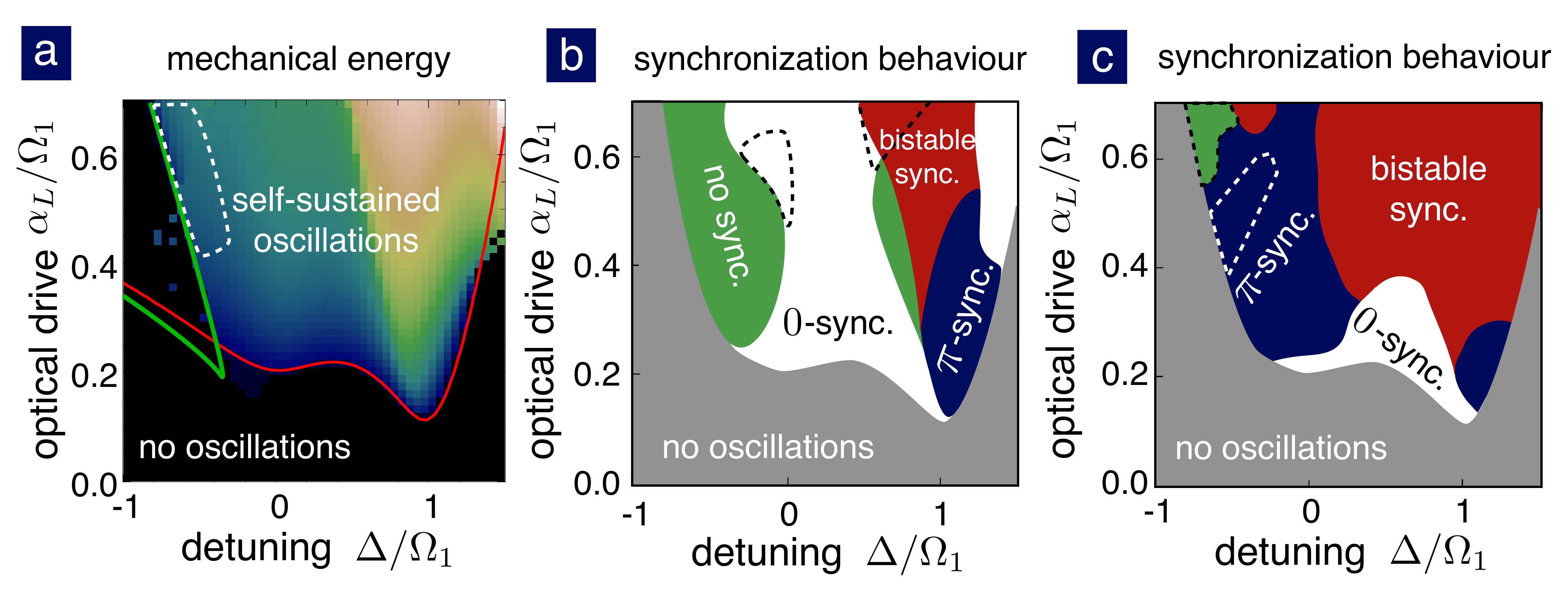}
\par\end{centering}

\protect\caption{Threshold of self-sustained oscillations and synchronization in the
absence of noise. (a) shows the mechanical energy (related to the
limit-cycle amplitude) of an optomechanical system as a function of
the laser drive $\alpha_{L}$ and the detuning $\Delta$. The red
line indicates the threshold of self-oscillations, $\Gamma+\Gamma_{\text{opt}}=0$.
The green line separates a region of optical multi-stability, the
dashed white line indicates a region where the self-oscillations show
strong amplitude modulation. (b) and (c) show schematic pictures of the classical, noiseless synchronization regimes of two coupled optomechanical systems. 
In (b) we use a small mechanical coupling $K/\Omega_{1}=0.05$ and different mechanical resonance frequencies, $\delta\Omega=0.075$ to give an overview
about all possible synchronization regimes. In (c) we use  $K/\Omega_{1}=0.15$ and $\delta\Omega=0$ instead, which significantly changes the phase diagram.
The dashed lines mark regions with even more complex behaviour, not necessarily showing synchronization. {[}Parameters: mechanical damping $\Gamma/\Omega_{1}=0.015$,
optical damping $\kappa/\Omega_{1}=0.3$, optomechanical coupling
$g_{0}/\Omega_{1}=0.3${]} \label{fig_Threshold}}
\end{figure}
We now introduce the model that we investigate throughout the rest
of this manuscript. We study two optomechanical systems which are
mechanically coupled, see Fig.~\ref{fig_Setup}(a). Our goal is to
analyze the synchronization behaviour in the presence of quantum and
thermal noise. Each optomechanical system consists of a driven optical
mode ($\hat{a}_{j}$) coupled to a mechanical mode ($\hat{b}_{j}$)
via radiation pressure. In a frame rotating at the laser drive frequency
$\omega_{L}$, the Hamiltonian of each optomechanical system is 

\begin{equation}
\hat{H}_{j}=-\hbar\Delta\hat{a}_{j}^{\dagger}\hat{a}_{j}+\hbar\Omega\hat{b}_{j}^{\dagger}\hat{b}_{j}-\hbar g_{0}\hat{a}_{j}^{\dagger}\hat{a}_{j}\left(\hat{b}_{j}+\hat{b}_{j}^{\dagger}\right)+\hbar\alpha_{L}(\hat{a}_{j}^{\dagger}+\hat{a}_{j}).
\end{equation}
Here, $\Delta=\omega_{L}-\omega_{c}$ denotes the detuning from the
cavity resonance $\omega_{c}$, $\Omega$ is the resonance frequency
of the mechanical mode, $g_{0}$ denotes the optomechanical single
photon coupling strength, and $\alpha_{L}$ is the laser driving strength.
Note that the optical and the mechanical systems experience damping
at a rate $\kappa$ and $\Gamma$, respectively, and these are described
by adding to the Hamiltonian a system-bath coupling in the usual manner:
$\hat{H}_{j}^{({\rm full)}}=\hat{H}_{j}+\hat{H}_{j}^{({\rm diss})}$. 

Driving a single optomechanical system with a blue-detuned laser causes
self-oscillations of the mechanical resonator when the optomechanically
induced negative damping $\Gamma_{\text{opt}}$ is larger than the
intrinsic damping of the oscillator $\Gamma$ \cite{2006_FM_DynamicalMultistability}.
In Fig.~\ref{fig_Threshold}(a) we show this threshold of self-oscillations. Throughout this work we only consider parameters such that the single optomechanical systems are above this threshold.
Self-oscillations at $\Delta<0$ occur due to the static optomechanical
shift which leads to an effective blue detuning, i.e.~$\Delta_{\text{eff}}>0$.
These limit-cycle oscillations, in the absence of noise, can effectively
be described by a fixed amplitude and a phase and are treated as a prerequisite
for synchronization throughout this work. We consider two self-oscillating optomechanical
systems that are coupled mechanically with strength $K$, such that
the total Hamiltonian of the system (except for the dissipative part)
reads

\begin{equation}
\hat{H}_{\text{tot}}=\sum_{j=1,2}\hat{H}_{j}-\hbar K\left(\hat{b}_{1}+\hat{b}_{1}^{\dagger}\right)\left(\hat{b}_{2}+\hat{b}_{2}^{\dagger}\right).\label{eq_FullHamiltonian}
\end{equation}
Experimentally, this coupling between the mechanical oscillators can
also be mediated by an optical coupling, cf.~Fig.~\ref{fig_Setup}(b)-(d).
In recent experiments \cite{2012_ZhangLipson_SynchronizationPRL,2013_BagheriTang_Synchronization,2015_Lipson_ArraySync},
a single joint optical mode was employed to couple the mechanical
oscillators. However, this mode served a dual purpose in creating
the limit cycles via a blue-detuned drive and providing the coupling.
Using an additional, independently driven optical mode for the coupling
would allow to tune the effective (optically induced) coupling independently
from the laser drive used to create the limit cycles. 

In experiments, the typical mode of operation is to have the two optomechanical
oscillators at slightly different intrinsic mechanical frequencies.
These start out un-synchronized but can synchronize upon changing
some parameter (e.g.~the laser drive strength, the detuning, or potentially
the coupling). This is schematically shown in Fig.~\ref{fig_Threshold}(b),
where we indicate the synchronization regimes in the absence of noise.
In accordance to the Hopf-Kuramoto model, cf.~Sec.~\ref{sec_CLsync},
there are unsynchronized regions and three different synchronization
regimes: (i) $0$\nobreakdash-synchronization, (ii) $\pi$-synchronization,
and (iii) classical bistable synchronization where the type of synchronization
depends on the initial conditions. Note that for different intrinsic
mechanical frequencies, $\delta\Omega=\Omega_{2}-\Omega_{1}\neq0$,
the relative phase $\delta\phi$ is not exactly $0$ or $\pi$ but
only close to one of these values and varies within the synchronization
regime. 

However, in the presence of noise it is already interesting to investigate
the behaviour even for identical frequencies. In particular, for large-scale
optomechanical arrays of identical oscillators, it has been found
that there is a synchronization transition as a function of noise
strength \cite{2013_Ludwig}. Moreover, in the present article we
will focus on noise-induced transitions between various synchronization
states. The observation of this physics does not rely on whether there
is an actual synchronization transition at lower values of the coupling.
Therefore, in most of our analysis, we will focus on identical systems,
i.e.~we assume all the parameters to be equal in both systems. 
In Fig.~\ref{fig_Threshold}(c) we schematically show the synchronization regimes for identical optomechanical systems and for a larger mechanical coupling 
$K$ than in Fig.~\ref{fig_Threshold}(b), but still in the absence of noise. It is important to note that both the mechanical detuning $\delta\Omega$ and the coupling $K$ have an influence on this  diagram: Not synchronized regions can become synchronized and synchronization regime borders are shifted.

We will comment on the dynamics of two coupled optomechanical oscillators
with different frequencies in Sec.~\ref{sec_SyncRegimes}.

\section{Methods\label{sec_Methods}}

In this work we use Langevin equations and quantum jump trajectories to study the system described by Hamiltonian (\ref{eq_FullHamiltonian}) in the presence of (quantum) noise. Here we want to briefly present both approaches and discuss their respective advantages and problems. 

Most of our results are computed with semi-classical Langevin equations. They are obtained by first deriving
quantum Langevin equations from \textcolor{black}{Hamiltonian~(\ref{eq_FullHamiltonian})
using input-output theory \cite{2008_WallsMilburn_QuantumOptics}.
We then adopt a semi-classical approach by turning the quantum Langevin
equations into classical Langevin equations for the complex amplitudes}
$\alpha_{j}$ and $\beta_{j}$, where the noise terms mimic the quantum-mechanical
zero-point fluctuations. This can be understood as a variant of the
``truncated Wigner approximation''. The semi-classical equations are:

\begin{equation}
\begin{alignedat}{1}\dot{\alpha}_{1}= & \left(i\Delta-\frac{\kappa}{2}\right)\alpha_{1}+ig_{0}\alpha_{1}\left(\beta_{1}+\beta_{1}^{*}\right)-i\alpha_{L}-\sqrt{\kappa}\alpha_{1\text{in}},\\
\dot{\beta}_{1}= & -\left(i\Omega+\frac{\Gamma}{2}\right)\beta_{1}+ig_{0}\left|\alpha_{1}\right|^{2}+iK\beta_{2}-\sqrt{\Gamma}\beta_{1\text{in}}.
\end{alignedat}
\label{eq_Leq}
\end{equation}
The corresponding equations for the second optomechanical system can
be obtained from Eqs.~(\ref{eq_Leq}) by exchanging the indices $1\longleftrightarrow2$.
Here, $\alpha_{j{\rm in}}(t)$ and $\beta_{j\text{in}}(t)$ represent
the optical and mechanical input noise, given by Gaussian stochastic
processes. Since complex numbers commute, they obviously cannot correctly
fulfill the input-output quantum noise correlators, $\langle\hat{a}_{j\text{in}}^{\dagger}(t)\hat{a}_{j\text{in}}(t')\rangle=0$
and $\langle\hat{a}_{j\text{in}}(t)\hat{a}_{j\text{in}}^{\dagger}(t')\rangle=\delta(t-t')$.
Instead, $\alpha_{j\text{in}}$ and $\beta_{j\text{in}}$ are made
to mimic quantum noise by fulfilling $\langle\alpha_{j\text{in}}(t)\alpha_{j\text{in}}^{*}(t')\rangle=\langle\alpha_{j\text{in}}^{*}(t)\alpha_{j\text{in}}(t')\rangle=\delta(t-t')/2$
(and likewise for $\beta_{j{\rm in}}$ for $T=0$). This approach allows to study also large parameter ranges with relatively low computational effort. 

Deep in the quantum regime it is initially not clear that these semi-classical Langevin equations describe the correct physical behaviour.  In order to verify the qualitative effects observed with Langevin equations, we also present a few results obtained with quantum jump trajectories \cite{1993_Molmer_quantumJumps,1998_Plenio_quantumJumps}, i.e.~an ``unraveling'' of the Lindblad master equation. Applying this method, the fully quantum system is simulated on an appropriately truncated Hilbert space. Notably, it
allows to work with wave functions, in contrast to the formalism of
the Lindblad master equation which requires the use of density matrices.
Hence, simulations of larger Hilbert spaces become feasible. In addition,
quantum jump trajectories give access to additional observables, for
instance the full counting statistics. For these reasons, quantum
jump trajectories have been applied in many different contexts. In
the field of quantum synchronization, such methods have been used
to study synchronization of qubits \cite{2009_QSync_Qubits}. In the
field of cavity optomechanics, they have been employed for instance
to discuss QND measurements, photon statistics and single photon optomechanics
\cite{2012_ML_EnhancedQuNonlinearities,2013_Kronwald_PhotonStatistics,2013_Akram_QJapplication,2014_Mirza_SinglePhotonSpectra_QJ}.
Another motivation is the recent experimental detection of individual
phonons in optomechanical systems \cite{2015_Cohen_phononCounting_Painter}
- a step towards monitoring full quantum jump trajectories in experiments.

To explain this approach \cite{1993_Molmer_quantumJumps,1998_Plenio_quantumJumps}, let us consider for a moment photon decay
in cavity 1. The ``unraveling'' discussed here corresponds to the
physical setup of placing a single photon detector at the output port
of the cavity. At each time step $\delta t$, the probability of a
single photon to leak out of the cavity (at temperature $T=0$) is
given by \textcolor{black}{$p=\kappa\delta t\langle\hat{a}_{1}^{\dagger}\hat{a}_{1}\rangle$}.
In the case of a photon loss it is detected at the output port and
the wave function is updated $\left|\Psi(t+\delta t)\right\rangle =\hat{a}_{1}\left|\Psi(t)\right\rangle $.
If no photon was lost to the environment, the wave function evolves
in time according to a (non-Hermitian) Hamiltonian that is obtained
by adding a term $-i\hbar(\kappa/2)\hat{a}_{1}^{\dagger}\hat{a}_{1}$.
This additional term accounts for the information gained about the
system by not observing a photon \cite{1990_Ueda_QJ_nonHermititanEvol}.
In both cases the state $\left|\Psi(t+\delta t)\right\rangle $ has
to be normalized before proceeding to the next time step. The treatment
of photon loss in cavity 2 and of phonon losses works analogously.
This simulation approach naturally accounts for quantum fluctuations. 

Although it would be favourable to use quantum jump trajectories throughout the whole study, this approach is only computationally feasible whenever the needed Hilbert space remains sufficiently small.  This leads to severe limitations in the choice of parameters and especially gives no access to the full classical-to-quantum crossover that is studied in Sec.~\ref{sec_QtoCL}. In Ref.~\cite{2008_LudwigMarquardt_OptomechInstab} it has been shown for a single optomechanical system that semi-classical Langevin equations produce good agreement with the full quantum theory. A systematic comparison for two coupled optomechanical systems is not possible, since the number of required Hilbert space dimensions is squared as compared to the single optomechanical system. We comment on this in Sec.~\ref{subsec_syncQNstrength} (cf.~Fig.~\ref{fig_Compare}(a) and the discussion below).

Note that quantum jump trajectories serve here as a numerical approach only. In experiments, single photon and phonon detection is not necessary. Instead, the mechanical oscillators could be measured using standard homodyning techniques. 

\section{Main Results\label{sec_Results}}

In this section we briefly state our main results, which will be discussed
in the following sections. Most notably, it is known even from the
classical theory that two coupled optomechanical oscillators can be
in either one of two synchronization states (with a phase difference
near $0$ or near $\pi$). We find a regime of ``mixed'' synchronization,
where transitions between $0$- and $\pi$-synchronization occur (Sec.~\ref{sec_FullQ}).
These transitions are driven by (quantum or thermal) noise and cannot be found
in the classical, noiseless situation. The average residence times
in the two synchronization states can differ and their ratio varies
with the system parameters (Sec.~\ref{sec_QtoCL}). Investigating
the classical-to-quantum transition, we find that mixed synchronization
can evolve from two different regimes in the classical, noiseless limit:
(i) there are already two stable synchronization states but in the
absence of noise there are no transitions, (ii) there is only one
stable synchronization state and only the presence of noise leads
to a second stable solution. Although the first sections are devoted to the investigation of quantum noise effects, we note that we find similar effects for thermal noise acting on the mechanical resonator. However, quantitative
differences remain due to the different nature of the noise source
(Sec.~\ref{sec_ThNoise}). We find that quantum noise effects should
dominate over thermal noise effects if the optomechanical cooperativity
is sufficiently large, and a large value of $g_{0}$ is not necessarily
required.

\section{Multistable Quantum Synchronization\label{sec_FullQ}}

First, we start by analyzing two coupled optomechanical systems, cf.~Fig.~\ref{fig_Setup}(a),
deep in the quantum regime. Quantum jump trajectories are used to initially investigate the full quantum dynamics. In the following, we consider a small-amplitude limit cycle and a
large single-photon coupling strength, $g_{0}/\kappa=1$. This ensures
that quantum fluctuations can potentially have a large impact on the
system's dynamics. Furthermore, small photon and phonon numbers are
necessary to keep the numerical simulations tractable, since they
determine the size of the truncated Hilbert space.

\begin{figure}
\centering{}\includegraphics[scale=0.35]{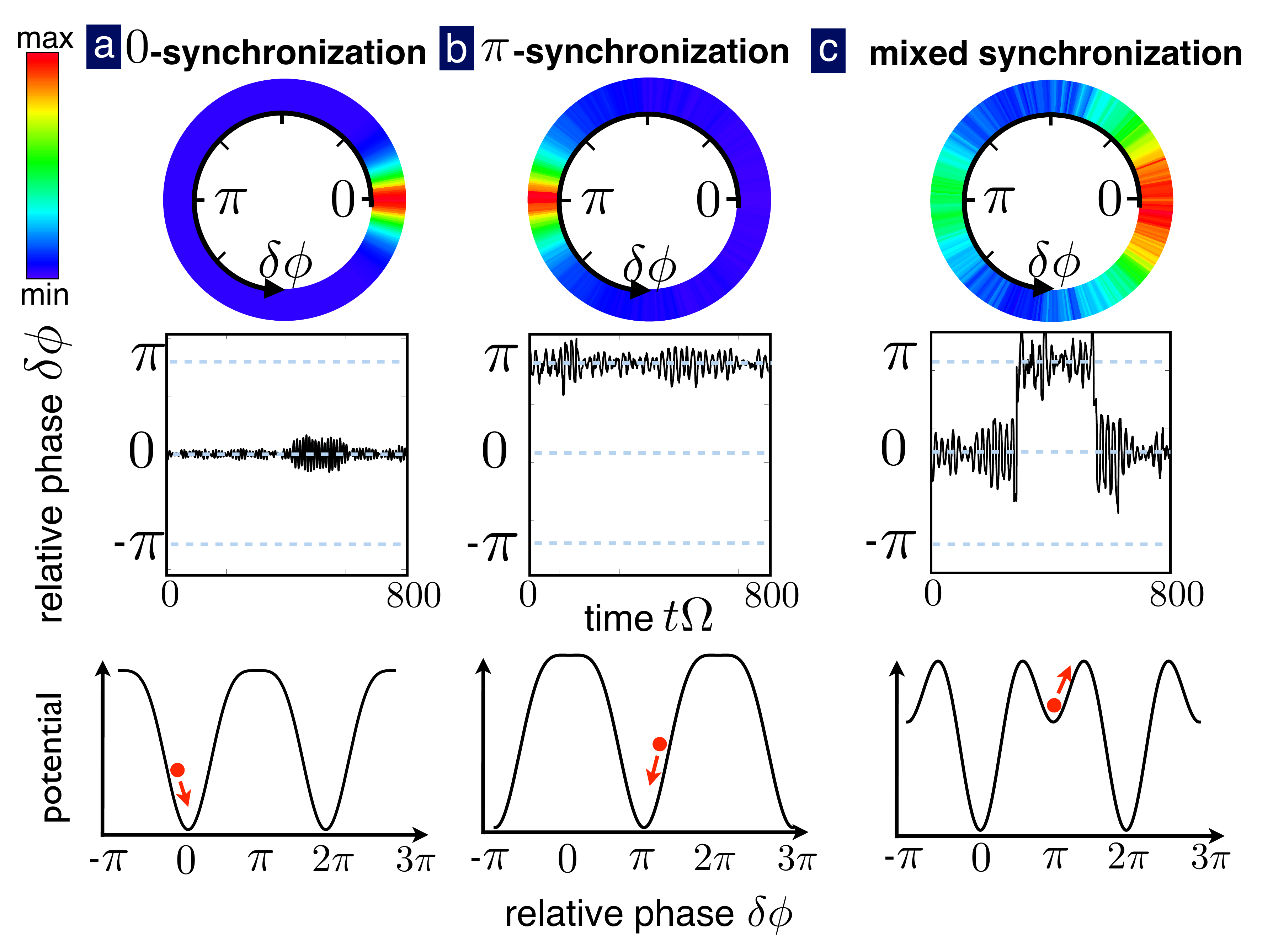}\protect\caption{Multistability in optomechanical quantum synchronization. Distribution
of the relative phase $\delta\phi$ in different synchronization regimes,
a typical sample of a corresponding quantum jump trajectory starting
in the steady state, and a sketch of the corresponding effective potential.
(a) shows $0$-synchronization, (b) shows $\pi$-synchronization,
and (c) shows mixed synchronization. A rotating wave approximation
for the mechanical coupling has been used\textcolor{blue}{.} {[}Parameters:
(a) mechanical coupling $K/\Omega=0.3$, mechanical damping $\Gamma/\Omega=0.015$;
(b) $K/\Omega=0.15$ , $\Gamma/\Omega=0.01$; (c) $K/\Omega=0.15$,
$\Gamma/\Omega=0.015$; other parameters are: optical damping $\kappa/\Omega=0.3$,
laser driving strength $\alpha_{L}/\Omega=0.3$, optomechanical coupling
$g_{0}/\kappa=1$, and optical detuning $\Delta/\Omega=0.15$. {]}\label{fig_PhaseStatAndTraj}}
\end{figure}
\textcolor{black}{To study synchronization we focus on the relative
phase $\delta\phi=\phi_{2}-\phi_{1}$ between the optomechanical oscillators.
Classically, $\delta\phi$ allows to identify synchronization ($\delta\phi=\text{const}$)
and distinguish the different synchronization regimes ($0$- and $\pi$-synchronization).
To extract the relative phase at each time step of a quantum jump
trajectory, we use that the motion of the uncoupled self-oscillators
in the absence of noise can effectively be described by} $\langle\hat{b}_{j}\rangle(t)\approx|b|e^{-i(\Omega t+\phi_{0j})}$,
with an initial random phase $\phi_{0j}$. Thus,\textcolor{black}{{}
the correlator }$\langle\hat{b}_{1}^{\dagger}\hat{b}_{2}\rangle\sim e^{-i\delta\phi}$
can be interpreted as a measure for the relative phase $\delta\phi$.
\textcolor{black}{In Fig.~\ref{fig_PhaseStatAndTraj} we show the
probability density of the relative phase obtained from the distribution
of $\delta\phi$ and a typical sample of the corresponding phase trajectory.
We find a $0$-synchronized regime, Fig.~\ref{fig_PhaseStatAndTraj}(a),
where the relative phase is predominantly close to $\delta\phi=0$.
Similarly, we find a $\pi$-synchronized regime, Fig.~\ref{fig_PhaseStatAndTraj}(b),
for different parameters.}\textcolor{blue}{{} }\textcolor{black}{As
expected, noise prevents perfect synchronization and thus the corresponding
maximum in the probability density has a finite width. Likewise, the
trajectories show fluctuations around either $\delta\phi=0$ or $\delta\phi=\pi$.
These two synchronization regimes have an analog in the classical,
noiseless limit, cf.~Sec.~\ref{sec_CLsync}. In addition, we find
another regime where the probability density of the phase difference
has maxima close to both $\delta\phi=0$ and $\delta\phi=\pi$, see
Fig.~\ref{fig_PhaseStatAndTraj}(c). The corresponding trajectory
shows that in this regime transitions between the two synchronization
states occur. We call this regime ``mixed'' synchronization. This
is in contrast to the classical, noiseless result, where the system
does not change its synchronization state with time, not even in the
regime of bistable synchronization. }

Similar to the classical, noiseless system \cite{2011_Heinrich_CollectiveDynamics}
we can understand these results in terms of an effective potential
for the relative phase $\delta\phi$. This is illustrated in the bottom
row of Fig.~\textcolor{black}{\ref{fig_PhaseStatAndTraj}. It offers
an intuitive understanding of our results. The noise makes the phase
fluctuate around the stable point(s) near $\delta\phi=0$ (and)or
$\delta\phi=\pi$.} In the case of mixed synchronization the effective
potential has two minima, near $\delta\phi=0$ and $\delta\phi=\pi$,
and quantum noise drives transitions between those two states. The
probability to be in either the $0$- or $\pi$-synchronized state
is given by the area of the corresponding maximum in the probability
density. It is associated to the depth of the minimum in the effective
phase potential. The ratio between the probabilities of the two states
can assume arbitrary values, depending on the parameters of the system.
Note that the absence of two distinct peaks in the probability density
does not necessarily mean that the system explores the region around
one synchronization state only. In fact, sometimes the trajectories
themselves can reveal short stretches of phase dynamics in the vicinity
of the other state. Nevertheless, if the fraction of time spent in
that other state is short, a second peak will not be visible.

\textcolor{black}{From this interpretation in terms of an effective
potential, it is clear that }the classical bistable synchronization
regime where two potential minima already exist (but no transitions
can occur), turns into a mixed synchronization regime (showing transitions)
in the presence of noise. However, our analysis in the following section
reveals that mixed synchronization can also appear for parameters
where classically there is only a single stable synchronization state.\textcolor{black}{}

\section{Classical-to-Quantum Crossover\label{sec_QtoCL}}

At the moment, the single-photon coupling strength $g_{0}$ is still
comparatively small in almost all experiments. As a consequence, quantum
effects have only been observed in the linearized regime where only
Gaussian states are produced. In the most promising cases \cite{2012_Chan_OptimizedOptomechanicalCrystalCavity,2011_Teufel_SidebandCooling_Nature},
$g_{0}/\kappa$ can take values up to $10^{-2}$ and $g_{0}/\Omega$
up to above $10^{-4}$. Much larger values have been reported for
experiments with cold atoms (up to around $g_{0}/\kappa\sim1$) \cite{2008_Murch_QuMeasurementBackaction},
but these do not operate in the ``good cavity limit'', i.e.~one
has $\kappa\gg\Omega$ in those experiments, precluding the observation
of single-photon strong coupling effects. Nevertheless, as experiments
are approaching the single-photon strong coupling regime, they will
gradually see increasingly strong effects of quantum fluctuations
even in non-linear dynamics. In this section, it is our aim to explore
the crossover between the classical regime (small $g_{0}/\kappa$)
and the quantum regime (large $g_{0}/\kappa$) with respect to optomechanical
quantum synchronization. We will focus our investigations mostly on
the mixed synchronization regime which is the most interesting one,
as we can have noise-induced transitions between the synchronization
states. In this section, we will disregard thermal noise, i.e.~we
assume temperature $T=0$, such that only quantum noise is present.
The ``classical'' regime we are discussing here is therefore the
noiseless limit of the classical equations of motion. We will later
remark on the effects of thermal noise (Sec.~\ref{sec_ThNoise}).

To explore the classical-to-quantum crossover, we want to effectively vary $\hbar$ while making sure to keep all the classical predictions unchanged. Notably, the optomechanical coupling strength $g_0=\partial \omega_c/\partial x$ depends on $\hbar$. For the simplest case of a Fabry-Pérot cavity of length $L$ with one static and one movable mirror the optomechanical coupling is $g_0=\omega_c(0)x_\text{ZPF}/L\sim\sqrt{\hbar}$, where $x_\text{ZPF}=\sqrt{\hbar/2 m \Omega}$ denotes the zero-point fluctuations of the mechanical oscillator of mass $m$. As discussed in Ref.~\cite{2008_LudwigMarquardt_OptomechInstab}, the ``quantum parameter'' $g_0/\kappa\sim\sqrt{\hbar}$ can thus be varied to effectively change the quantum noise strength. This implies that all classical ($\hbar$-independent) parameters
\textcolor{black}{($\kappa/\Omega$, $\Gamma/\Omega$, $\Delta/\Omega$,
$K/\Omega$, $g_{0}\alpha_{L}/\Omega^{2}$)} are kept fixed while
$g_{0}$ is modified. To see that the quantum parameter has indeed this anticipated effect, it is very helpful to rescale the amplitudes $\alpha_{j}$ and $\beta_{j}$, Eqs.~(\ref{eq_Leq}), 
such that they tend to a well-defined finite value in the classical
limit $g_{0}/\kappa\rightarrow0$ \cite{2008_LudwigMarquardt_OptomechInstab}.
This can be achieved by defining $\tilde{\alpha}_{j}=g_{0}\alpha_{j}$
and $\tilde{\beta}_{j}=g_{0}\beta_{j}$.  While $\left|\alpha_{1}\right|^{2}$
gives the number of photons (a ``quantum-mechanical'' quantity),
the rescaled version $\left|\tilde{\alpha}_{1}\right|^{2}=g_{0}^{2}\left|\alpha_{1}\right|^{2}\sim\hbar\left|\alpha_{1}\right|^{2}$
is proportional to the energy $\hbar\omega_{c}\left|\alpha_{1}\right|^{2}$
inside the cavity (i.e.~a classical quantity). A similar argument
applies to $\tilde{\beta}_{j}$. With this rescaling, we find the
following equations:

\begin{equation}
\begin{alignedat}{1}\dot{\tilde{\alpha}}_{1}= & \left(i\Delta-\frac{\kappa}{2}\right)\tilde{\alpha}_{1}+i\tilde{\alpha}_{1}\left(\tilde{\beta}_{1}+\tilde{\beta}_{1}^{*}\right)-i\tilde{\alpha}_{L}-\sqrt{\kappa}\tilde{\alpha}_{1\text{in}},\\
\dot{\tilde{\beta}}_{1}= & -\left(i\Omega+\frac{\Gamma}{2}\right)\tilde{\beta}_{1}+i\left|\tilde{\alpha}_{1}\right|^{2}+iK\tilde{\beta}_{2}-\sqrt{\Gamma}\tilde{\beta}_{1\text{in}}.
\end{alignedat}
\label{eq_Leq-1}
\end{equation}
Here $\tilde{\alpha}_{L}=g_{0}\alpha_{L}$ is the rescaled laser-driving
amplitude that we keep fixed while varying $g_{0}$. The
important observation here is that $g_{0}$ has been completely eliminated
from the equations and now only appears in the strength of the quantum
noise: we now have $\langle\tilde{\alpha}_{j\text{in}}(t)\tilde{\alpha}_{j\text{in}}^{*}(t')\rangle=g_{0}^{2}\delta(t-t')/2$
(and likewise for $\tilde{\beta}_{j\text{in}}$) which indeed vanishes in the classical limit of $g_{0}\sim\sqrt{\hbar}\rightarrow0$. 

If we consider mechanical
thermal noise at finite temperature $T$ (as discussed in Sec.~\ref{sec_ThNoise}),
we have $\langle\tilde{\beta}_{j\text{in}}(t)\tilde{\beta}_{j\text{in}}^{*}(t')\rangle=g_{0}^{2}(n_{{\rm th}}+1/2)\delta(t-t')$.
Here, $n_{\text{th}}$ denotes the thermal occupancy of the bath coupled
to the mechanical oscillator. The product $g_{0}^{2}n_{{\rm th}}\sim\hbar n_{{\rm th}}$
becomes independent of $\hbar$ in the classical limit $k_{B}T\gg\hbar\Omega$,
where $\hbar n_{{\rm th}}\approx k_{B}T/\Omega$ and $k_{B}$ is Boltzmann's
constant. In summary, the rescaled equations (\ref{eq_Leq-1}) nicely
show how an increase of $g_{0}$ can indeed be viewed solely as an
increase of the strength of ``quantum noise'' in our system.

Note that if $g_{0}\rightarrow 0$ we have to
increase the laser driving strength $\alpha_{L}$ to keep $\tilde{\alpha}_{L}=g_{0}\alpha_{L}$
constant which means that the total light energy circulating inside
the optical cavity is constant. Due to $\hbar\rightarrow0$, this corresponds to an increasing
number of photons inside the cavity and thus increases drastically
the size of the Hilbert space necessary for the full quantum simulation. Therefore,
quantum jump trajectories are not suitable to explore the full quantum-to-classical
crossover and we have to apply Langevin equations. 

\subsection{Synchronization as a Function of Quantum Noise Strength\label{subsec_syncQNstrength}}

\begin{figure}
\centering{}\includegraphics[scale=0.4]{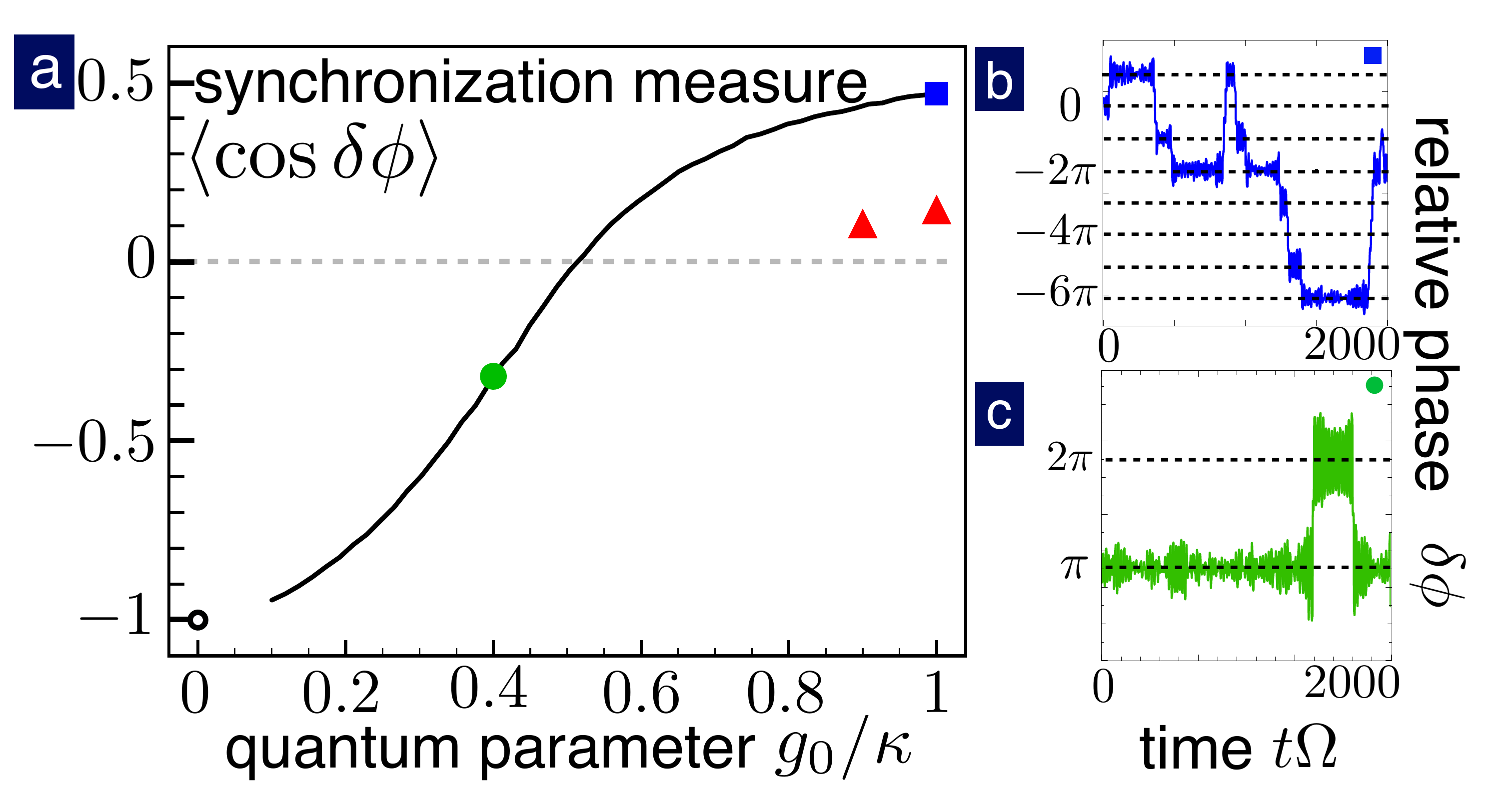}\protect\caption{Quantum-to-classical crossover. (a) Synchronization measure as a function
of the quantum parameter $g_{0}/\kappa$, using Langevin equations
(black line), classical (noiseless) Langevin equations (black circle,
$g_{0}/\kappa=0$), and with quantum jumps (red triangles).\textcolor{blue}{{}
}(b) and (c) show trajectories of the relative phase in the quantum
regime $g_{0}/\kappa=1$ (blue square) and for $g_{0}/\kappa=0.4$
(green dot). Other parameters are as in Fig.~\textcolor{blue}{\ref{fig_PhaseStatAndTraj}}(c),
a rotating wave approximation has been used for the mechanical coupling\textcolor{blue}{.}\label{fig_Compare}}
\end{figure}
In all studies of synchronization, one needs to select suitable quantities
that measure the degree of synchronization. In this context, it is
important to note that for any \textit{finite} noisy system (subject
to quantum and/or thermal noise), there is no sharp synchronization
transition and correspondingly no unambiguous measure that displays
nonanalytic behaviour at any parameter value. Before proceeding to
our results, we summarize and discuss the synchronization measures
adopted here which have to be combined to obtain a full picture: (i)
the probability density of $\delta\phi$, (ii) the average phase factor
$\left\langle e^{-i\delta\phi}\right\rangle $, and (iii) individual
trajectories.

The probability density of the relative phase $\delta\phi$ is either
mostly flat (no synchronization) or, as shown in Fig.~\ref{fig_PhaseStatAndTraj},
centered predominantly around $0$ or $\pi$, or it may have two peaks,
according to the synchronization regime. In the following, we aim
to compress the information contained in the phase distribution into
one quantity and calculate the normalized correlator $C=\langle\hat{b}_{1}^{\dagger}\hat{b}_{2}\rangle/\sqrt{\langle\hat{b}_{1}^{\dagger}\hat{b}_{1}\rangle\langle\hat{b}_{2}^{\dagger}\hat{b}_{2}\rangle}\approx\left\langle e^{-i\delta\phi}\right\rangle $.
Its real value, $\text{{Re}}[C]=\langle\cos\delta\phi\rangle$, distinguishes
the three different synchronization regimes: (i) $\langle\cos\delta\phi\rangle\approx1$
for $0$-synchronization, (ii) $\langle\cos\delta\phi\rangle\approx-1$
for $\pi$-synchronization, (iii) intermediate values of $\langle\cos\delta\phi\rangle$
for mixed synchronization. However, this measure has its limitations:
When $\delta\phi$ is more or less evenly distributed (no synchronization)
$\langle\cos\delta\phi\rangle\approx0$, this cannot be distinguished
from a mixed synchronization situation where almost equal time is
spent in the $0$- and $\pi$-synchronized states. Furthermore, even
in the absence of synchronization (and even in the noiseless case)
the phase $\delta\phi$ may spend an increased amount of time around
certain values. This leads to a finite value of $\langle\cos\delta\phi\rangle$,
and similarly would also show up in the phase distribution. A solution
to this problem is to simultaneously look at a part of the corresponding
trajectory, where synchronization can easily be distinguished from
an unsynchronized state. Instead, one could start to use more complicated
correlators, e.g.~$\langle\hat{b}_{1}^{\dagger}\hat{b}_{1}^{\dagger}\hat{b}_{2}\hat{b}_{2}\rangle\sim e^{-i2\delta\phi}$.
This correlator allows to distinguish unsynchronized states from synchronized
states, but an additional measure is needed to distinguish $0$- from
$\pi$-synchronization. Note that the imaginary part of the above
defined correlator, $\text{Im}[C]=\langle\sin\delta\phi\rangle$,
can be used as well. However, in the special case of identical optomechanical
systems $\text{Im}[C]\approx0$ due to the symmetry of the system. 

In Fig.~\ref{fig_Compare}(a) we show how $\langle\cos\delta\phi\rangle$
varies as a function of the quantum parameter $g_{0}/\kappa$. We
chose parameters that lead to a mixed synchronization regime for larger
values of $g_{0}/\kappa$. In the deep quantum regime ($g_{0}/\kappa\rightarrow1)$
the probability $P_{0}$ to find the system in the $0$-synchronized
state is larger than the probability $P_{\pi}$ to find the phase
around $\pi$, such that $\langle\cos\delta\phi\rangle>0$. Going
towards smaller values of $g_{0}/\kappa$, the ratio $P_{\pi}/P_{0}$
increases, such that eventually $\langle\cos\delta\phi\rangle<0$.
Finally, we should reach the classical (noiseless) limit, when $g_{0}/\kappa\rightarrow0$.
It turns out that, for the parameters adopted here, the classical
solution always ends up in the $\pi$-synchronized state, independent
of initial conditions. This implies that there is only one minimum
in the effective potential. We conclude that the system has turned
from a mixed synchronization regime into a purely $\pi$-synchronized
regime as the quantum parameter was reduced. This cannot be understood
in the simple picture of a noise-independent phase potential. We will
discuss this kind of behaviour in more detail later on (Sec.~\ref{sub:LargeQuantumParameter}).

Note that the
calculations for Fig.~\ref{fig_Compare} have been performed using Langevin equations; although
in the deep quantum regime, two data points were also acquired with
quantum jump trajectories (red triangles). They are shifted as compared to the Langevin
results, but show the same trend. 
We expect that the difference between the Langevin and quantum jump results decreases for smaller quantum parameters $g_0/\kappa$, as it is the case for a single optomechanical system \cite{2008_LudwigMarquardt_OptomechInstab}. Since smaller $g_0/\kappa$ require a significantly larger Hilbert space for the quantum jump simulations, we cannot compute this for our coupled system. For large quantum parameter $g_0/\kappa\sim1$, qualitative differences between the full quantum model and the Langevin equations have been already observed in Ref.~\cite{2008_LudwigMarquardt_OptomechInstab} as well. Especially a shift of the detuning $\Delta$ was reported, that could be determined numerically also for our system. Taking this detuning shift into account would improve the agreement of our results, although differences remain. Here, we show the uncorrected outcomes of both approaches.

\subsection{Residence Times in the Mixed Synchronization Regime}

\begin{figure}
\centering{}\includegraphics[scale=0.35]{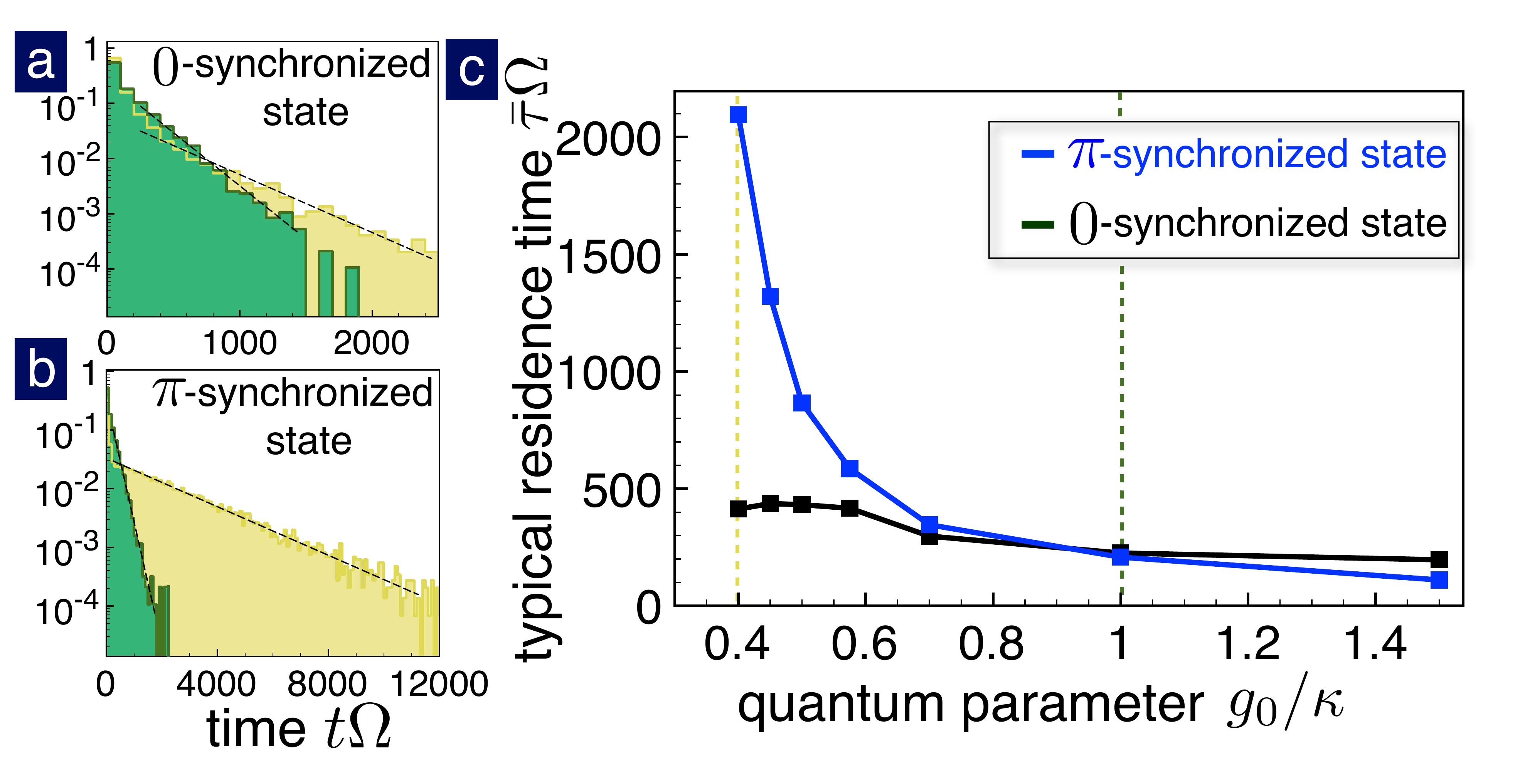}\protect\caption{Residence times of synchronization states. (a) and (b) show the distribution
of residence times in the $0$-synchronized and $\pi$-synchronized
state. The yellow histogram shows the distribution for $g_{0}/\kappa=0.4$,
the green histogram gives the result deeper in the quantum regime,
$g_{0}/\kappa=1$, for comparison. Note the different scale of the
time axis in (a) and (b). The corresponding typical time scale $\tau_{0}$
or $\tau_{\pi}$ is obtained from a fit and is shown in (c) as a function
of the quantum parameter $g_{0}/\kappa$. Results are obtained using
Langevin equations, with the following parameters: optical damping
$\kappa/\Omega=0.3$, mechanical damping $\Gamma/\Omega=0.015$, detuning
$\Delta/\Omega=-1/30$, mechanical coupling $K/\Omega=0.15$, $g_{0}\alpha_{L}/\Omega^{2}=0.09=\text{const}$,
and a rotating wave approximation for the mechanical coupling term.
\label{fig_HistTimes} }
\end{figure}
The measure $\left\langle \cos\delta\phi\right\rangle $ quantifies
the fraction of total time spent in $0$-synchronized parts as compared
to $\pi$-synchronized parts. However, it does not provide any information
about the rate of transitions between the two synchronization states.
Based on the effective potential picture, one might expect the transition
rates to be determined by the barrier height and the noise strength.
In particular, for larger effective noise strengths $g_{0}/\kappa$,
we expect more frequent transitions. This behaviour is qualitatively
visible in Fig.~\ref{fig_Compare}(b) and (c). However, as concluded
in the previous section, the potential picture is not sufficient to
explain all observations. Thus, we now turn to a quantitative analysis
and discuss how the transition rates between $0$- and $\pi$-synchronized
states (i.e.~the typical residence times $\bar{\tau}$) change during
the classical-to-quantum crossover. We extract the fluctuating residence
times from the phase trajectories and obtain their distribution. The
results are shown in Fig.~\ref{fig_HistTimes}(a) and (b) for the
$0$- and $\pi$-synchronized states, for two different quantum parameters
$g_{0}/\kappa$. In all cases the probability densities decay exponentially
with time, $\sim e^{-\tau/\bar{\tau}}$, and the average residence
time $\bar{\tau}$ is obtained from a fit to the distribution. The
extracted average residence times $\tau_{0}$ and $\tau_{\pi}$ for
the two states are shown in Fig.~\ref{fig_HistTimes}(c) as a function
of the quantum parameter. Note that the ratio of residence times equals
the ratio of probabilities, $\tau_{0}/\tau_{\pi}=P_{0}/P_{\pi}$.
Nevertheless, the dependence of the times $\tau_{0},\,\tau_{\pi}$
on $g_{0}/\kappa$ reveals new information.

We have chosen parameters such that at $g_{0}/\kappa=1$ \textcolor{black}{both
$0$-synchronized and $\pi$-synchronized parts have almost equal
average residence times. This corresponds to $\langle\cos\delta\phi\rangle\approx0$
and $P_{0}\approx P_{\pi}$. Furthermore, in the classical limit $g_{0}/\kappa=0$
the system is $\pi$-synchronized only. When the classical limit $g_{0}/\kappa\rightarrow0$
is approached, we find that both $\tau_{0}$ and $\tau_{\pi}$ increase.
As expected, $\tau_{\pi}$ increases much faster than $\tau_{0}$
and eventually diverges for $g_{0}/\kappa\rightarrow0$, as the system
gets trapped forever in the $\pi$-state. In contrast, $\tau_{0}$
increases first when decreasing $g_{0}/\kappa$, but then saturates
at a finite level. }Such a behaviour is unexpected based on the simple
phase potential picture, where a fixed potential would imply diverging
residence times for both states in the noiseless limit. The behaviour
observed here hinges on the fact that the synchronization regime switches
from ``mixed'' to ``$\pi$'' as one reduces the quantum parameter
(i.e.~reduces the quantum noise). The observations would change significantly
for different parameters, where the system always stays in the mixed
regime, for any value $g_{0}/\kappa>0$. Then, one expects the simple
picture of a fixed phase potential to be approximately correct and
both residence times to diverge as the noise is becoming weaker.

\subsection{Noise-induced Synchronization Bistability}

\label{sub:LargeQuantumParameter}

\begin{figure}
\centering{}\includegraphics[scale=0.35]{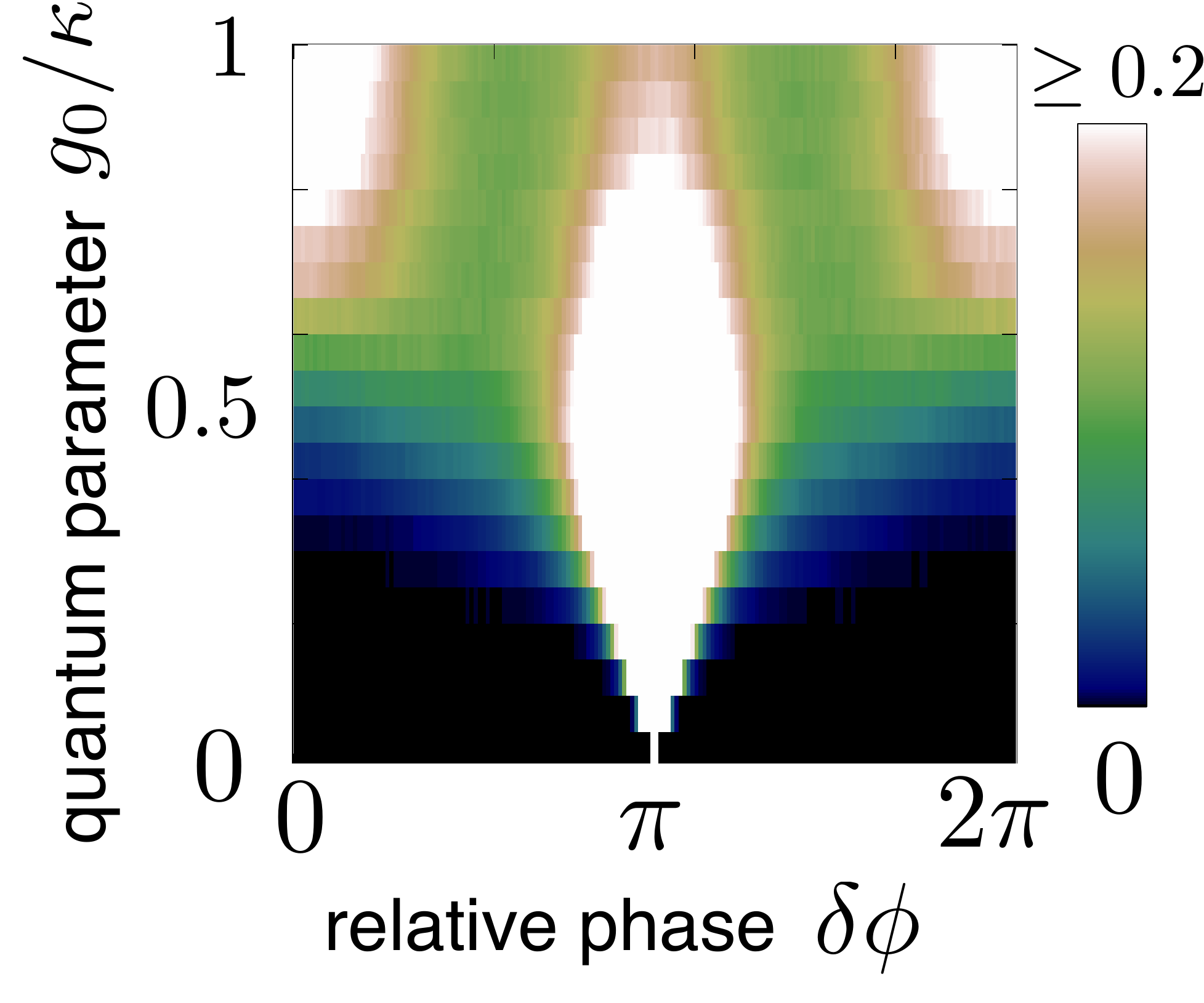}\protect\caption{\textcolor{black}{Appearance of bistability. The distribution of the
relative phase as a function of the quantum parameter $g_{0}/\kappa$.
P}arameters are as in Fig.~\ref{fig_HistTimes}, but with mechanical
coupling $K/\Omega=0.1$. \textcolor{blue}{\label{fig_bistab_qVSnth}}}
\end{figure}
In the previous sections we have explained that some basic features
of the classical-to-quantum crossover, like the increase of the residence
times with decreasing quantum parameter, can be understood as effects
of a decreasing quantum noise strength. The decrease of noise strength
leads to less frequent transitions across an energy barrier in the
effective potential. However, we made also less easily explained observations:
(i) the reverse in the order of $\tau_{0}$ and $\tau_{\pi}$ as $g_{0}$
is being reduced, (ii) the saturation of $\tau_{0}$ at low noise
levels, (iii) the disappearance of stable $0$-synchronization (for
the applied parameters) in the classical limit $g_{0}/\kappa=0$.
Whereas (i) and (ii) could originate from more complicated potential
shapes with a combination of broad and narrow minima, (iii) suggests
that the effective potential itself changes when the quantum parameter
is varied. In Fig.~\ref{fig_bistab_qVSnth} we show how the distribution
of $\delta\phi$ evolves as a function of $g_{0}/\kappa$. For very
small values of $g_{0}/\kappa$, there is only a single peak close
to $\delta\phi\approx\pi$, in accordance with the single stable solution
of the classical limit. While increasing the quantum parameter, this
peak is first broadened. The increasing quantum noise strength allows
the system to explore more of the effective potential around the minimum.
A significant accumulation close to $\delta\phi\approx0$ appears
only for rather large quantum parameters, signaling the appearance
of a second stable solution, i.e.~a second minimum in the effective
potential at $\delta\phi\approx0$. 

In addition to the above described appearance of a second stable solution,
there are also parameter regions where already in the classical regime
the effective potential has minima at both $\delta\phi=0$ and $\delta\phi=\pi$
(classical bistable synchronization). In this case quantum noise naturally
drives transitions between the two synchronization states as soon
as it is added to the description of the system. The number of observed
transitions then naturally depends on both the noise strength as well
as the potential shape.

\section{Overview of synchronization regimes\label{sec_SyncRegimes}}

\textcolor{black}{In the previous sections we have shown examples
of the different synchronization regimes in the presence of quantum
noise and studied the properties of mixed synchronization in more
detail. In the following, we map out the different synchronization
regimes as a function of the mechanical coupling $K$ and the quantum
parameter $g_{0}/\kappa$. Furthermore, we also discuss the case of
detuned mechanical oscillators. }

\begin{figure*}[t]
\centering{}\includegraphics[scale=0.35]{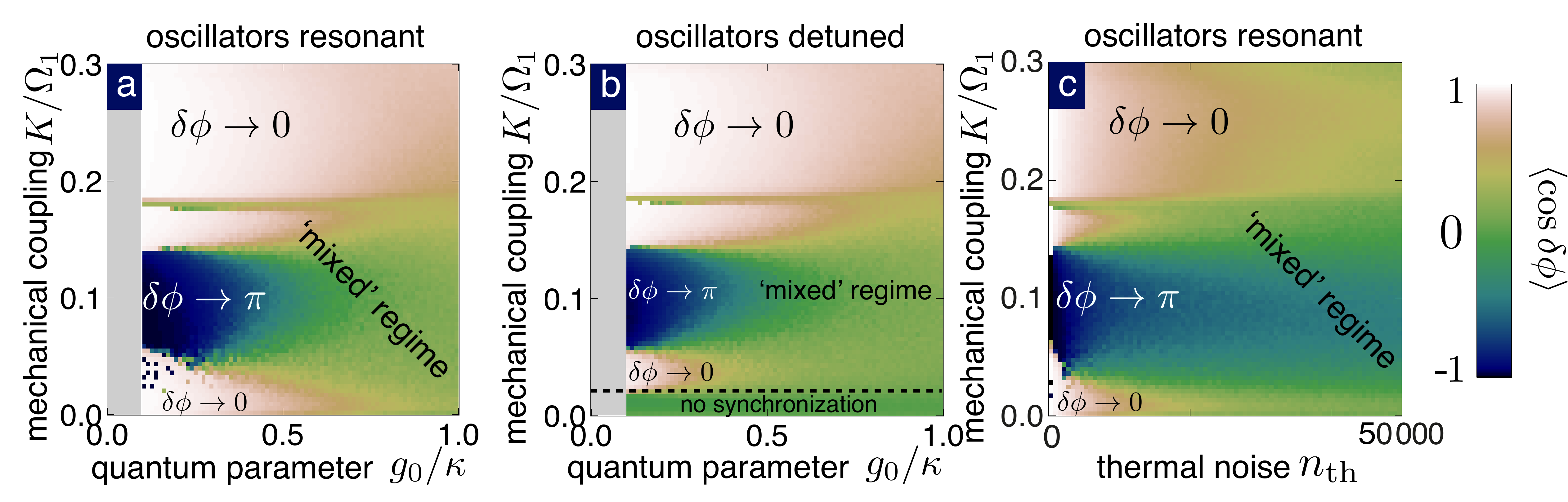}\protect\caption{Synchronization regimes versus mechanical coupling and noise. The
measure $\langle\cos\delta\phi\rangle$ as a function of the quantum
parameter $g_{0}/\kappa$ (i.e.~effective quantum noise strength)
and of the mechanical coupling strength $K/\Omega_{1}$, for identical
optomechanical systems (a) and for systems with mechanical frequency
detuning $\delta\Omega/\Omega_{1}=(\Omega_{2}-\Omega_{1})/\Omega_{1}=0.05$
(b). In (c), we display the effects of thermal noise $n_{\text{th}}$.
In that case, we show $\langle\cos\delta\phi\rangle$ as a function
of coupling $K$ and environmental thermal phonon number $n_{\text{th}}$,
with a constant $g_{0}/\kappa=0.01$ and for zero $\delta\Omega$.
The grey region in (a) and (b) was not simulated. The dashed line
in (b) is the synchronization threshold as observed in the classical,
noiseless limit $g_{0}/\kappa=0$. Parameters are as in Fig.~\ref{fig_HistTimes}.
\label{fig_quToClassOverview} }
\end{figure*}

Figure \ref{fig_quToClassOverview}(a) shows the
synchronization measure $\langle\cos\delta\phi\rangle$ for resonant
oscillators. We have indicated the synchronization regimes (which
are not sharply delineated). For the applied parameters, we find $0$-synchronization
for large mechanical coupling $K$ in both the quantum and classical
regime. In the classical, noiseless limits $\langle\cos\delta\phi\rangle\rightarrow1$,
indicating less fluctuations around the synchronization state. In
contrast, at smaller mechanical coupling, we find more complicated
behaviour: there is mixed synchronization for $g_{0}/\kappa\sim1$,
while the classical limit $g_{0}/\kappa\rightarrow0$ selects either
$0-$ or $\pi-$synchronization, depending on the mechanical coupling
$K$. The ``pixelated'' region in Fig.~\ref{fig_quToClassOverview}(a)
indicates that the system is multistable even in the classical limit.
There, the residence times have become so large that the system is
stuck in a random synchronization state depending on initial conditions
and the transient behaviour. 
Notably, the closer the system is to a border of synchronization regimes in the classical limit (this can be seen in Fig.~\ref{fig_quToClassOverview}(a) for small $g_0/\kappa$ when $K$ is varied), the smaller the noise strength ($g_0/\kappa$) that is needed to lead to mixed synchronization. As an example,  in the "middle" of the classically $\pi$-synchronized regime (at about $K/\Omega_1\approx0.1$) similar mixed synchronization as compared to the system close to the regime border ($K/\Omega_1\lesssim0.15$) appears only for larger values of the quantum parameter. An exception is of course the classically bistable regime, where mixed synchronization appears naturally as soon as there is noise.
An interesting feature appears close
to $K/\Omega\approx0.18$, where the measure $\langle\cos\delta\phi\rangle$
shows a sharp dip in the middle of a $0$-synchronized region. We
suspect a non-linear resonance, since the oscillator trajectory $x_{j}(t)$
is no longer simply sinusoidal and period-doubling is observed. At
the same time, the oscillation amplitude increases. For larger values
of the mechanical coupling $K$, the trajectories are simply sinusoidal
again, with the same frequency as for coupling strengths below the
feature. 

Up to now, we have studied the ideal case of identical mechanical
oscillators. We now turn to the case where the mechanical oscillators
have slightly different resonance frequencies, i.e.~\textcolor{black}{$\delta\Omega=\Omega_{2}-\Omega_{1}\neq0$.
This is a typical situation in experiments, since fabrication inaccuracies
lead to deviations between two systems. Figure \ref{fig_quToClassOverview}(b)
shows how $\delta\Omega$ affects synchronization. A finite $\delta\Omega$
corresponds to a tilted effective potential in the classical, noiseless
limit, where a finite threshold for synchronization appears \cite{2011_Heinrich_CollectiveDynamics}.
This classical threshold is indicated in Fig.~\ref{fig_quToClassOverview}(b)
with a dashed line. Similar behaviour is visible for quantum parameters
$g_{0}/\kappa>0$. However, for large $g_{0}/\kappa$ it is not possible
to determine the onset of synchronization, because the measure $\langle\cos\delta\phi\rangle$
cannot distinguish between no synchronization and mixed synchronization.
Instead, we now also have to analyze the trajectories in more detail,
which reveal mixed synchronization for sufficiently large mechanical
coupling. A significant deviation from the classical threshold cannot
be observed at the given resolution. We expect the threshold to slightly
increase for larger quantum parameter $g_{0}/\kappa$. At the same
time, however, the threshold is also smeared out due to quantum noise.}

Here, we have chosen to show the dependence of the synchronization
regimes on the mechanical coupling strength $K$ and the quantum parameter
$g_{0}/\kappa$. Note that other parameters influence the synchronization
type as well. In Fig.~\ref{fig_Threshold} we have already seen that the laser driving strength $\alpha_L$ and the detuning $\Delta$ influence the classical synchronization regimes. Also the mechanical damping $\Gamma$ can affect the observed synchronization, cf.~Fig.~\ref{fig_PhaseStatAndTraj}b and c. However, first of all it already influences the limit cycles of individual optomechanical systems by modifying the threshold to self-sustained oscillations. In addition, $\Gamma$ has an influence on the mechanical noise strength. Note that, when changing
these parameters, care has to be taken to remain on a stable limit
cycle for each optomechanical system. 

\section{Thermal noise\label{sec_ThNoise}}

\textcolor{black}{So far, we assumed zero temperature environments
for both the optical and the mechanical mode. In this section we investigate
the effects of thermal mechanical noise on synchronization.}

For our study we use the Langevin equations (\ref{eq_Leq}) with modified
mechanical noise correlators to account for the coupling of the mechanical
oscillators to a finite temperature bath, $\langle\beta_{j\text{in}}(t)\beta_{j\text{in}}^{*}(t')\rangle=\langle\beta_{j\text{in}}^{*}(t)\beta_{j\text{in}}(t')\rangle=(n_{\text{th}}+1/2)\delta(t-t')$,
where $n_{\text{th}}$ denotes the thermal occupancy of the bath.
Hence, both quantum and thermal noise are included. For optical frequencies
in the visible spectrum the effective thermal occupation of the optical
bath is very small. Thus, the assumption of an optical bath at zero
temperature is valid and the optical input noise terms are not modified. 

In Fig.~\ref{fig_quToClassOverview}(c) we show an overview of the
synchronization regimes as a function of thermal noise $n_{\text{th}}$.
Here, we chose a comparatively small quantum parameter $g_{0}/\kappa=0.01$
in order to observe mainly effects due to thermal noise. For small
$n_{\text{th}}$ the results are similar to Fig.~\ref{fig_quToClassOverview}(a)
at small $g_{0}/\kappa$. In both cases the influence of the quantum
or thermal noise is still weak. For increasing thermal noise strength
$n_{{\rm th}}$, we find qualitatively the same behaviour as for increasing
quantum noise. However, quantitative differences appear with increasing
$n_{\text{th}}$: Even though a mixed synchronization regime can develop
for both quantum and thermal noise, the evolution of the relative
weight of both synchronization states is different.

In the following, we want to estimate the critical thermal noise strength
$n_{\text{th}}^{*}$ at which thermal and quantum noise should have
a comparable effect. At lower temperatures, quantum noise will dominate.
The main source of quantum fluctuations is the laser shot noise (for
the parameters explored here). Thus, we estimate the effect of optical
quantum fluctuations on the mechanical oscillator, with the (symmetrized)
shot-noise spectrum evaluated at the mechanical resonance frequency
$\Omega$ \cite{2007_FM_SidebandCooling,2014_Review},

\begin{equation}
S_{FF}^{\text{SN}}(\Omega)=\frac{1}{2}\left(\frac{\hbar g_{0}}{x_{\text{ZPF}}}\right)^{2}\langle\hat{a}_{j}^{\dagger}\hat{a}_{j}\rangle\left(\frac{\kappa}{(\kappa/2)^{2}+(\Omega+\Delta)^{2}}\left.+\frac{\kappa}{(\kappa/2)^{2}+(\Omega-\Delta)^{2}}\right).\right.\label{eq_SFF}
\end{equation}
We expect
similar effects from quantum and thermal noise if the shot-noise spectrum
at the mechanical resonance frequency $\Omega$ becomes equal to the
thermal force spectrum, $S_{FF}^{\text{SN}}(\Omega)=S_{FF}^{\text{th}}$.
The thermal noise spectrum at temperatures $k_{B}T\gg\hbar\Omega$
is 
\begin{equation}
S_{FF}^{\text{th}}=2m\Gamma k_{B}T,
\end{equation}
where $T$ is the temperature of the thermal mechanical bath. Setting
$S_{FF}^{{\rm SN}}$ and $S_{FF}^{{\rm th}}$ equal, we find (in the
resolved sideband regime $\kappa\ll\Omega$ at $\Delta\approx\Omega$):
\begin{equation}
n_{\text{th}}^{*}=k_{B}T/\hbar\Omega=\mathcal{C}/2,\label{eq_critNth}
\end{equation}
where we used the optomechanical cooperativity \cite{2014_Review}
\begin{equation}
\mathcal{C}=\frac{4g_{0}^{2}}{\kappa\Gamma}\langle\hat{a}_{j}^{\dagger}\hat{a}_{j}\rangle.
\end{equation}
This approach suggests that observing quantum noise phenomena does
not necessarily require a large $g_{0}/\kappa$ and very low temperatures.
Instead, if the cooperativity is sufficiently large (comparable to
values that enable ground-state cooling, $\mathcal{C}>n_{\text{th}}$),
quantum noise should dominate the behaviour of the system even in
the presence of thermal noise. 

However, depending on parameters, we find large deviations from this
simple expectation. In these cases, the real shot noise spectrum is
no longer well described by the weak-coupling expression of $S_{FF}^{\text{SN}}(\Omega)$
given above, and the actual noise strength may have a much larger
value. Consequently, the transition between behaviour dominated by
quantum noise vs.~that dominated by thermal noise takes place at
much larger values of $n_{{\rm th}}^{*}$ than those predicted by
Eq.~(\ref{eq_critNth}). In other words, it should be even easier
to observe quantum noise in optomechanical synchronization than the
naive ansatz would lead one to expect.

Here, we don't observe that quantum noise can be exactly mapped to
thermal noise. This is already evident in the rescaled Langevin equations
(\ref{eq_Leq-1}). Physically, the thermal force spectrum acting on
the mechanical oscillator is flat in frequency, whereas the optical
shot-noise spectrum is frequency dependent and is also modified by
the dynamics of the system.

\section{Conclusions}

We have investigated the effects of quantum and thermal noise on two
coupled optomechanical limit-cycle oscillators. One usually expects
that noise prevents strict synchronization, i.e.~exact phase locking
and a sharp transition to synchronization. Here we have shown that
fluctuations additionally drive transitions between $0$- and $\pi$-synchronization,
i.e.~the two synchronization states that can appear in the absence
of noise. We have discussed the residence times of these states and
observed a smooth crossover between different synchronization regimes.
Finally, we have compared the effects of quantum and thermal noise.
We have argued that it should be possible to experimentally reach
the regime where quantum noise dominates. This should happen when
the optomechanical cooperativity is large enough for ground state
cooling.

For further investigations it would be useful to identify a measure
that can genuinely distinguish between an unsynchronized regime and
$0$-, $\pi$- and mixed synchronization. Finally, it will be very
interesting to extend the insights obtained here to large optomechanical
arrays.
\begin{acknowledgments}
This work was supported by the ITN cQOM as well as ERC OPTOMECH. We
thank Steven Habraken and Roland Lauter for discussions and Stefan
Walter for his critical reading of the manuscript.
\end{acknowledgments}

\bibliographystyle{apsrev4-1}

\begin{thebibliography}{74}%
\makeatletter
\providecommand \@ifxundefined [1]{%
 \@ifx{#1\undefined}
}%
\providecommand \@ifnum [1]{%
 \ifnum #1\expandafter \@firstoftwo
 \else \expandafter \@secondoftwo
 \fi
}%
\providecommand \@ifx [1]{%
 \ifx #1\expandafter \@firstoftwo
 \else \expandafter \@secondoftwo
 \fi
}%
\providecommand \natexlab [1]{#1}%
\providecommand \enquote  [1]{``#1''}%
\providecommand \bibnamefont  [1]{#1}%
\providecommand \bibfnamefont [1]{#1}%
\providecommand \citenamefont [1]{#1}%
\providecommand \href@noop [0]{\@secondoftwo}%
\providecommand \href [0]{\begingroup \@sanitize@url \@href}%
\providecommand \@href[1]{\@@startlink{#1}\@@href}%
\providecommand \@@href[1]{\endgroup#1\@@endlink}%
\providecommand \@sanitize@url [0]{\catcode `\\12\catcode `\$12\catcode
  `\&12\catcode `\#12\catcode `\^12\catcode `\_12\catcode `\%12\relax}%
\providecommand \@@startlink[1]{}%
\providecommand \@@endlink[0]{}%
\providecommand \url  [0]{\begingroup\@sanitize@url \@url }%
\providecommand \@url [1]{\endgroup\@href {#1}{\urlprefix }}%
\providecommand \urlprefix  [0]{URL }%
\providecommand \Eprint [0]{\href }%
\providecommand \doibase [0]{http://dx.doi.org/}%
\providecommand \selectlanguage [0]{\@gobble}%
\providecommand \bibinfo  [0]{\@secondoftwo}%
\providecommand \bibfield  [0]{\@secondoftwo}%
\providecommand \translation [1]{[#1]}%
\providecommand \BibitemOpen [0]{}%
\providecommand \bibitemStop [0]{}%
\providecommand \bibitemNoStop [0]{.\EOS\space}%
\providecommand \EOS [0]{\spacefactor3000\relax}%
\providecommand \BibitemShut  [1]{\csname bibitem#1\endcsname}%
\let\auto@bib@innerbib\@empty
%</preamble>
\bibitem [{\citenamefont {Aspelmeyer}\ \emph {et~al.}(2014)\citenamefont
  {Aspelmeyer}, \citenamefont {Kippenberg},\ and\ \citenamefont
  {Marquardt}}]{2014_Review}%
  \BibitemOpen
  \bibfield  {author} {\bibinfo {author} {\bibfnamefont {M.}~\bibnamefont
  {Aspelmeyer}}, \bibinfo {author} {\bibfnamefont {T.~J.}\ \bibnamefont
  {Kippenberg}}, \ and\ \bibinfo {author} {\bibfnamefont {F.}~\bibnamefont
  {Marquardt}},\ }\href {\doibase 10.1103/RevModPhys.86.1391} {\bibfield
  {journal} {\bibinfo  {journal} {Rev. Mod. Phys.}\ }\textbf {\bibinfo {volume}
  {86}},\ \bibinfo {pages} {1391} (\bibinfo {year} {2014})}\BibitemShut
  {NoStop}%
\bibitem [{\citenamefont {Hill}\ \emph {et~al.}(2012)\citenamefont {Hill},
  \citenamefont {Safavi-Naeini}, \citenamefont {Chan},\ and\ \citenamefont
  {Painter}}]{2012_Hill_CoherentWavelengthConv}%
  \BibitemOpen
  \bibfield  {author} {\bibinfo {author} {\bibfnamefont {J.~T.}\ \bibnamefont
  {Hill}}, \bibinfo {author} {\bibfnamefont {A.~H.}\ \bibnamefont
  {Safavi-Naeini}}, \bibinfo {author} {\bibfnamefont {J.}~\bibnamefont {Chan}},
  \ and\ \bibinfo {author} {\bibfnamefont {O.}~\bibnamefont {Painter}},\ }\href
  {\doibase 10.1038/ncomms2201} {\bibfield  {journal} {\bibinfo  {journal}
  {Nat. Commun.}\ }\textbf {\bibinfo {volume} {3}},\ \bibinfo {pages} {1196}
  (\bibinfo {year} {2012})}\BibitemShut {NoStop}%
\bibitem [{\citenamefont {Dong}\ \emph {et~al.}(2012)\citenamefont {Dong},
  \citenamefont {Fiore}, \citenamefont {Kuzyk},\ and\ \citenamefont
  {Wang}}]{2012_Dong_OMDarkMode}%
  \BibitemOpen
  \bibfield  {author} {\bibinfo {author} {\bibfnamefont {C.}~\bibnamefont
  {Dong}}, \bibinfo {author} {\bibfnamefont {V.}~\bibnamefont {Fiore}},
  \bibinfo {author} {\bibfnamefont {M.~C.}\ \bibnamefont {Kuzyk}}, \ and\
  \bibinfo {author} {\bibfnamefont {H.}~\bibnamefont {Wang}},\ }\href
  {http://www.sciencemag.org/content/338/6114/1609} {\bibfield  {journal}
  {\bibinfo  {journal} {Science}\ }\textbf {\bibinfo {volume} {338}},\ \bibinfo
  {pages} {1609} (\bibinfo {year} {2012})}\BibitemShut {NoStop}%
\bibitem [{\citenamefont {Grudinin}\ \emph {et~al.}(2010)\citenamefont
  {Grudinin}, \citenamefont {Lee}, \citenamefont {Painter},\ and\ \citenamefont
  {Vahala}}]{2010_Grudinin_PhononLaserAction}%
  \BibitemOpen
  \bibfield  {author} {\bibinfo {author} {\bibfnamefont {I.~S.}\ \bibnamefont
  {Grudinin}}, \bibinfo {author} {\bibfnamefont {H.}~\bibnamefont {Lee}},
  \bibinfo {author} {\bibfnamefont {O.}~\bibnamefont {Painter}}, \ and\
  \bibinfo {author} {\bibfnamefont {K.~J.}\ \bibnamefont {Vahala}},\ }\href
  {\doibase 10.1103/PhysRevLett.104.083901} {\bibfield  {journal} {\bibinfo
  {journal} {Phys. Rev. Lett.}\ }\textbf {\bibinfo {volume} {104}},\ \bibinfo
  {pages} {083901} (\bibinfo {year} {2010})}\BibitemShut {NoStop}%
\bibitem [{\citenamefont {Bahl}\ \emph {et~al.}(2012)\citenamefont {Bahl},
  \citenamefont {Tomes}, \citenamefont {Marquardt},\ and\ \citenamefont
  {Carmon}}]{2012_Bahl_SpontaneousBrillouinCooling}%
  \BibitemOpen
  \bibfield  {author} {\bibinfo {author} {\bibfnamefont {G.}~\bibnamefont
  {Bahl}}, \bibinfo {author} {\bibfnamefont {M.}~\bibnamefont {Tomes}},
  \bibinfo {author} {\bibfnamefont {F.}~\bibnamefont {Marquardt}}, \ and\
  \bibinfo {author} {\bibfnamefont {T.}~\bibnamefont {Carmon}},\ }\href
  {http://dx.doi.org/10.1038/nphys2206} {\bibfield  {journal} {\bibinfo
  {journal} {Nat. Phys.}\ }\textbf {\bibinfo {volume} {8}},\ \bibinfo {pages}
  {203} (\bibinfo {year} {2012})}\BibitemShut {NoStop}%
\bibitem [{\citenamefont {Wang}\ and\ \citenamefont
  {Clerk}(2012)}]{2012_Wang_QuantumStateTransfer}%
  \BibitemOpen
  \bibfield  {author} {\bibinfo {author} {\bibfnamefont {Y.-D.}\ \bibnamefont
  {Wang}}\ and\ \bibinfo {author} {\bibfnamefont {A.~A.}\ \bibnamefont
  {Clerk}},\ }\href {\doibase 10.1103/PhysRevLett.108.153603} {\bibfield
  {journal} {\bibinfo  {journal} {Phys. Rev. Lett.}\ }\textbf {\bibinfo
  {volume} {108}},\ \bibinfo {pages} {153603} (\bibinfo {year}
  {2012})}\BibitemShut {NoStop}%
\bibitem [{\citenamefont {Woolley}\ and\ \citenamefont
  {Clerk}(2014)}]{2014_Woolley_TwoModeSqueezing}%
  \BibitemOpen
  \bibfield  {author} {\bibinfo {author} {\bibfnamefont {M.~J.}\ \bibnamefont
  {Woolley}}\ and\ \bibinfo {author} {\bibfnamefont {A.~A.}\ \bibnamefont
  {Clerk}},\ }\href {\doibase 10.1103/PhysRevA.89.063805} {\bibfield  {journal}
  {\bibinfo  {journal} {Phys. Rev. A}\ }\textbf {\bibinfo {volume} {89}},\
  \bibinfo {pages} {063805} (\bibinfo {year} {2014})}\BibitemShut {NoStop}%
\bibitem [{\citenamefont {Woolley}\ and\ \citenamefont
  {Clerk}(2013)}]{2013_Woolley_BAEmsmt}%
  \BibitemOpen
  \bibfield  {author} {\bibinfo {author} {\bibfnamefont {M.~J.}\ \bibnamefont
  {Woolley}}\ and\ \bibinfo {author} {\bibfnamefont {A.~A.}\ \bibnamefont
  {Clerk}},\ }\href {\doibase 10.1103/PhysRevA.87.063846} {\bibfield  {journal}
  {\bibinfo  {journal} {Phys. Rev. A}\ }\textbf {\bibinfo {volume} {87}},\
  \bibinfo {pages} {063846} (\bibinfo {year} {2013})}\BibitemShut {NoStop}%
\bibitem [{\citenamefont {Paternostro}\ \emph {et~al.}(2007)\citenamefont
  {Paternostro}, \citenamefont {Vitali}, \citenamefont {Gigan}, \citenamefont
  {Kim}, \citenamefont {Brukner}, \citenamefont {Eisert},\ and\ \citenamefont
  {Aspelmeyer}}]{2007_Paternostro_EntanglementWithLight}%
  \BibitemOpen
  \bibfield  {author} {\bibinfo {author} {\bibfnamefont {M.}~\bibnamefont
  {Paternostro}}, \bibinfo {author} {\bibfnamefont {D.}~\bibnamefont {Vitali}},
  \bibinfo {author} {\bibfnamefont {S.}~\bibnamefont {Gigan}}, \bibinfo
  {author} {\bibfnamefont {M.~S.}\ \bibnamefont {Kim}}, \bibinfo {author}
  {\bibfnamefont {C.}~\bibnamefont {Brukner}}, \bibinfo {author} {\bibfnamefont
  {J.}~\bibnamefont {Eisert}}, \ and\ \bibinfo {author} {\bibfnamefont
  {M.}~\bibnamefont {Aspelmeyer}},\ }\href {\doibase
  10.1103/PhysRevLett.99.250401} {\bibfield  {journal} {\bibinfo  {journal}
  {Phys. Rev. Lett.}\ }\textbf {\bibinfo {volume} {99}},\ \bibinfo {eid}
  {250401} (\bibinfo {year} {2007})}\BibitemShut {NoStop}%
\bibitem [{\citenamefont {Deng}\ \emph {et~al.}(2014)\citenamefont {Deng},
  \citenamefont {Habraken},\ and\ \citenamefont
  {Marquardt}}]{2014_Deng_CV_entanglement}%
  \BibitemOpen
  \bibfield  {author} {\bibinfo {author} {\bibfnamefont {Z.~J.}\ \bibnamefont
  {Deng}}, \bibinfo {author} {\bibfnamefont {S.~J.~M.}\ \bibnamefont
  {Habraken}}, \ and\ \bibinfo {author} {\bibfnamefont {F.}~\bibnamefont
  {Marquardt}},\ }\href {http://arxiv.org/abs/1406.7815} {\bibfield  {journal}
  {\bibinfo  {journal} {arXiv:1406.7815}\ } (\bibinfo {year}
  {2014})}\BibitemShut {NoStop}%
\bibitem [{\citenamefont {Wang}\ \emph {et~al.}(2015)\citenamefont {Wang},
  \citenamefont {Chesi},\ and\ \citenamefont
  {Clerk}}]{2015_Wang_OutputEntanglement_3ModeOM}%
  \BibitemOpen
  \bibfield  {author} {\bibinfo {author} {\bibfnamefont {Y.-D.}\ \bibnamefont
  {Wang}}, \bibinfo {author} {\bibfnamefont {S.}~\bibnamefont {Chesi}}, \ and\
  \bibinfo {author} {\bibfnamefont {A.~A.}\ \bibnamefont {Clerk}},\ }\href
  {\doibase 10.1103/PhysRevA.91.013807} {\bibfield  {journal} {\bibinfo
  {journal} {Phys. Rev. A}\ }\textbf {\bibinfo {volume} {91}},\ \bibinfo
  {pages} {013807} (\bibinfo {year} {2015})}\BibitemShut {NoStop}%
\bibitem [{\citenamefont {Heinrich}\ \emph {et~al.}(2010)\citenamefont
  {Heinrich}, \citenamefont {Harris},\ and\ \citenamefont
  {Marquardt}}]{2010_Heinrich_PhotonShuttle}%
  \BibitemOpen
  \bibfield  {author} {\bibinfo {author} {\bibfnamefont {G.}~\bibnamefont
  {Heinrich}}, \bibinfo {author} {\bibfnamefont {J.~G.~E.}\ \bibnamefont
  {Harris}}, \ and\ \bibinfo {author} {\bibfnamefont {F.}~\bibnamefont
  {Marquardt}},\ }\href {\doibase 10.1103/PhysRevA.81.011801} {\bibfield
  {journal} {\bibinfo  {journal} {Phys. Rev. A}\ }\textbf {\bibinfo {volume}
  {81}},\ \bibinfo {pages} {011801} (\bibinfo {year} {2010})}\BibitemShut
  {NoStop}%
\bibitem [{\citenamefont {{Wu}}\ \emph {et~al.}(2013)\citenamefont {{Wu}},
  \citenamefont {{Heinrich}},\ and\ \citenamefont
  {{Marquardt}}}]{2013_Wu_LandauZenerPhononLasing}%
  \BibitemOpen
  \bibfield  {author} {\bibinfo {author} {\bibfnamefont {H.}~\bibnamefont
  {{Wu}}}, \bibinfo {author} {\bibfnamefont {G.}~\bibnamefont {{Heinrich}}}, \
  and\ \bibinfo {author} {\bibfnamefont {F.}~\bibnamefont {{Marquardt}}},\
  }\href {http://iopscience.iop.org/1367-2630/15/12/123022/} {\bibfield
  {journal} {\bibinfo  {journal} {New J. Phys.}\ }\textbf {\bibinfo {volume}
  {15}},\ \bibinfo {pages} {123022} (\bibinfo {year} {2013})}\BibitemShut
  {NoStop}%
\bibitem [{\citenamefont {Zhang}\ \emph {et~al.}(2012)\citenamefont {Zhang},
  \citenamefont {Wiederhecker}, \citenamefont {Manipatruni}, \citenamefont
  {Barnard}, \citenamefont {McEuen},\ and\ \citenamefont
  {Lipson}}]{2012_ZhangLipson_SynchronizationPRL}%
  \BibitemOpen
  \bibfield  {author} {\bibinfo {author} {\bibfnamefont {M.}~\bibnamefont
  {Zhang}}, \bibinfo {author} {\bibfnamefont {G.~S.}\ \bibnamefont
  {Wiederhecker}}, \bibinfo {author} {\bibfnamefont {S.}~\bibnamefont
  {Manipatruni}}, \bibinfo {author} {\bibfnamefont {A.}~\bibnamefont
  {Barnard}}, \bibinfo {author} {\bibfnamefont {P.}~\bibnamefont {McEuen}}, \
  and\ \bibinfo {author} {\bibfnamefont {M.}~\bibnamefont {Lipson}},\ }\href
  {\doibase 10.1103/PhysRevLett.109.233906} {\bibfield  {journal} {\bibinfo
  {journal} {Phys. Rev. Lett.}\ }\textbf {\bibinfo {volume} {109}},\ \bibinfo
  {pages} {233906} (\bibinfo {year} {2012})}\BibitemShut {NoStop}%
\bibitem [{\citenamefont {Maldovan}\ and\ \citenamefont
  {Thomas}(2006)}]{2006_Maldovan_OMcrystals}%
  \BibitemOpen
  \bibfield  {author} {\bibinfo {author} {\bibfnamefont {M.}~\bibnamefont
  {Maldovan}}\ and\ \bibinfo {author} {\bibfnamefont {E.~L.}\ \bibnamefont
  {Thomas}},\ }\href {http://dx.doi.org/10.1063/1.2216885} {\bibfield
  {journal} {\bibinfo  {journal} {Appl. Phys. Lett.}\ }\textbf {\bibinfo
  {volume} {88}},\ \bibinfo {pages} {251907} (\bibinfo {year}
  {2006})}\BibitemShut {NoStop}%
\bibitem [{\citenamefont {Eichenfield}\ \emph {et~al.}(2009)\citenamefont
  {Eichenfield}, \citenamefont {Chan}, \citenamefont {Camacho}, \citenamefont
  {Vahala},\ and\ \citenamefont
  {Painter}}]{2009_Eichenfield_OptomechanicalCrystals}%
  \BibitemOpen
  \bibfield  {author} {\bibinfo {author} {\bibfnamefont {M.}~\bibnamefont
  {Eichenfield}}, \bibinfo {author} {\bibfnamefont {J.}~\bibnamefont {Chan}},
  \bibinfo {author} {\bibfnamefont {R.~M.}\ \bibnamefont {Camacho}}, \bibinfo
  {author} {\bibfnamefont {K.~J.}\ \bibnamefont {Vahala}}, \ and\ \bibinfo
  {author} {\bibfnamefont {O.}~\bibnamefont {Painter}},\ }\href
  {http://dx.doi.org/10.1038/nature08524} {\bibfield  {journal} {\bibinfo
  {journal} {Nature}\ }\textbf {\bibinfo {volume} {462}},\ \bibinfo {pages}
  {78} (\bibinfo {year} {2009})}\BibitemShut {NoStop}%
\bibitem [{\citenamefont {Safavi-Naeini}\ \emph {et~al.}(2010)\citenamefont
  {Safavi-Naeini}, \citenamefont {Alegre}, \citenamefont {Winger},\ and\
  \citenamefont {Painter}}]{2010_Safavi-Naeini_Slotted2D}%
  \BibitemOpen
  \bibfield  {author} {\bibinfo {author} {\bibfnamefont {A.~H.}\ \bibnamefont
  {Safavi-Naeini}}, \bibinfo {author} {\bibfnamefont {T.~P.~M.}\ \bibnamefont
  {Alegre}}, \bibinfo {author} {\bibfnamefont {M.}~\bibnamefont {Winger}}, \
  and\ \bibinfo {author} {\bibfnamefont {O.}~\bibnamefont {Painter}},\ }\href
  {\doibase 10.1063/1.3507288} {\bibfield  {journal} {\bibinfo  {journal}
  {Appl. Phys. Lett.}\ }\textbf {\bibinfo {volume} {97}},\ \bibinfo {eid}
  {181106} (\bibinfo {year} {2010})}\BibitemShut {NoStop}%
\bibitem [{\citenamefont {Gavartin}\ \emph {et~al.}(2011)\citenamefont
  {Gavartin}, \citenamefont {Braive}, \citenamefont {Sagnes}, \citenamefont
  {Arcizet}, \citenamefont {Beveratos}, \citenamefont {Kippenberg},\ and\
  \citenamefont {Robert-Philip}}]{2011_Sagnes2DCrystaldefect}%
  \BibitemOpen
  \bibfield  {author} {\bibinfo {author} {\bibfnamefont {E.}~\bibnamefont
  {Gavartin}}, \bibinfo {author} {\bibfnamefont {R.}~\bibnamefont {Braive}},
  \bibinfo {author} {\bibfnamefont {I.}~\bibnamefont {Sagnes}}, \bibinfo
  {author} {\bibfnamefont {O.}~\bibnamefont {Arcizet}}, \bibinfo {author}
  {\bibfnamefont {A.}~\bibnamefont {Beveratos}}, \bibinfo {author}
  {\bibfnamefont {T.~J.}\ \bibnamefont {Kippenberg}}, \ and\ \bibinfo {author}
  {\bibfnamefont {I.}~\bibnamefont {Robert-Philip}},\ }\href {\doibase
  10.1103/PhysRevLett.106.203902} {\bibfield  {journal} {\bibinfo  {journal}
  {Phys. Rev. Lett.}\ }\textbf {\bibinfo {volume} {106}},\ \bibinfo {pages}
  {203902} (\bibinfo {year} {2011})}\BibitemShut {NoStop}%
\bibitem [{\citenamefont {Laude}\ \emph {et~al.}(2011)\citenamefont {Laude},
  \citenamefont {Beugnot}, \citenamefont {Benchabane}, \citenamefont {Pennec},
  \citenamefont {Djafari-Rouhani}, \citenamefont {Papanikolaou}, \citenamefont
  {Escalante},\ and\ \citenamefont {Martinez}}]{2011_Laude_phoxonicCrystal}%
  \BibitemOpen
  \bibfield  {author} {\bibinfo {author} {\bibfnamefont {V.}~\bibnamefont
  {Laude}}, \bibinfo {author} {\bibfnamefont {J.-C.}\ \bibnamefont {Beugnot}},
  \bibinfo {author} {\bibfnamefont {S.}~\bibnamefont {Benchabane}}, \bibinfo
  {author} {\bibfnamefont {Y.}~\bibnamefont {Pennec}}, \bibinfo {author}
  {\bibfnamefont {B.}~\bibnamefont {Djafari-Rouhani}}, \bibinfo {author}
  {\bibfnamefont {N.}~\bibnamefont {Papanikolaou}}, \bibinfo {author}
  {\bibfnamefont {J.~M.}\ \bibnamefont {Escalante}}, \ and\ \bibinfo {author}
  {\bibfnamefont {A.}~\bibnamefont {Martinez}},\ }\href {\doibase
  10.1364/OE.19.009690} {\bibfield  {journal} {\bibinfo  {journal} {Opt.
  Express}\ }\textbf {\bibinfo {volume} {19}},\ \bibinfo {pages} {9690}
  (\bibinfo {year} {2011})}\BibitemShut {NoStop}%
\bibitem [{\citenamefont {Safavi-Naeini}\ \emph {et~al.}(2014)\citenamefont
  {Safavi-Naeini}, \citenamefont {Hill}, \citenamefont {Meenehan},
  \citenamefont {Chan}, \citenamefont {Gr\"oblacher},\ and\ \citenamefont
  {Painter}}]{2014_Safavi-Naeini_OmCrystal}%
  \BibitemOpen
  \bibfield  {author} {\bibinfo {author} {\bibfnamefont {A.~H.}\ \bibnamefont
  {Safavi-Naeini}}, \bibinfo {author} {\bibfnamefont {J.~T.}\ \bibnamefont
  {Hill}}, \bibinfo {author} {\bibfnamefont {S.}~\bibnamefont {Meenehan}},
  \bibinfo {author} {\bibfnamefont {J.}~\bibnamefont {Chan}}, \bibinfo {author}
  {\bibfnamefont {S.}~\bibnamefont {Gr\"oblacher}}, \ and\ \bibinfo {author}
  {\bibfnamefont {O.}~\bibnamefont {Painter}},\ }\href {\doibase
  10.1103/PhysRevLett.112.153603} {\bibfield  {journal} {\bibinfo  {journal}
  {Phys. Rev. Lett.}\ }\textbf {\bibinfo {volume} {112}},\ \bibinfo {pages}
  {153603} (\bibinfo {year} {2014})}\BibitemShut {NoStop}%
\bibitem [{\citenamefont {Chang}\ \emph {et~al.}(2011)\citenamefont {Chang},
  \citenamefont {Safavi-Naeini}, \citenamefont {Hafezi},\ and\ \citenamefont
  {Painter}}]{2011_Chang_SlowingAndStoppingLight_NJP}%
  \BibitemOpen
  \bibfield  {author} {\bibinfo {author} {\bibfnamefont {D.~E.}\ \bibnamefont
  {Chang}}, \bibinfo {author} {\bibfnamefont {A.~H.}\ \bibnamefont
  {Safavi-Naeini}}, \bibinfo {author} {\bibfnamefont {M.}~\bibnamefont
  {Hafezi}}, \ and\ \bibinfo {author} {\bibfnamefont {O.}~\bibnamefont
  {Painter}},\ }\href {http://stacks.iop.org/1367-2630/13/i=2/a=023003}
  {\bibfield  {journal} {\bibinfo  {journal} {New J. Phys.}\ }\textbf {\bibinfo
  {volume} {13}},\ \bibinfo {pages} {023003} (\bibinfo {year}
  {2011})}\BibitemShut {NoStop}%
\bibitem [{\citenamefont {Schmidt}\ \emph
  {et~al.}(2015{\natexlab{a}})\citenamefont {Schmidt}, \citenamefont {Peano},\
  and\ \citenamefont {Marquardt}}]{2015_Schmidt_OMDiracPhysics}%
  \BibitemOpen
  \bibfield  {author} {\bibinfo {author} {\bibfnamefont {M.}~\bibnamefont
  {Schmidt}}, \bibinfo {author} {\bibfnamefont {V.}~\bibnamefont {Peano}}, \
  and\ \bibinfo {author} {\bibfnamefont {F.}~\bibnamefont {Marquardt}},\ }\href
  {http://iopscience.iop.org/1367-2630/17/2/023025/} {\bibfield  {journal}
  {\bibinfo  {journal} {New J. Phys.}\ }\textbf {\bibinfo {volume} {17}},\
  \bibinfo {pages} {023025} (\bibinfo {year} {2015}{\natexlab{a}})}\BibitemShut
  {NoStop}%
\bibitem [{\citenamefont {Tomadin}\ \emph {et~al.}(2012)\citenamefont
  {Tomadin}, \citenamefont {Diehl}, \citenamefont {Lukin}, \citenamefont
  {Rabl},\ and\ \citenamefont
  {Zoller}}]{2012_TomadinZoller_ReservoirEngineeringOptomechanicalArray}%
  \BibitemOpen
  \bibfield  {author} {\bibinfo {author} {\bibfnamefont {A.}~\bibnamefont
  {Tomadin}}, \bibinfo {author} {\bibfnamefont {S.}~\bibnamefont {Diehl}},
  \bibinfo {author} {\bibfnamefont {M.~D.}\ \bibnamefont {Lukin}}, \bibinfo
  {author} {\bibfnamefont {P.}~\bibnamefont {Rabl}}, \ and\ \bibinfo {author}
  {\bibfnamefont {P.}~\bibnamefont {Zoller}},\ }\href {\doibase
  10.1103/PhysRevA.86.033821} {\bibfield  {journal} {\bibinfo  {journal} {Phys.
  Rev. A}\ }\textbf {\bibinfo {volume} {86}},\ \bibinfo {pages} {033821}
  (\bibinfo {year} {2012})}\BibitemShut {NoStop}%
\bibitem [{\citenamefont {Schmidt}\ \emph
  {et~al.}(2015{\natexlab{b}})\citenamefont {Schmidt}, \citenamefont
  {Ke{\ss}ler}, \citenamefont {Peano}, \citenamefont {Painter},\ and\
  \citenamefont {Marquardt}}]{2015_Schmidt_MagneticFieldForPhotons}%
  \BibitemOpen
 \bibfield  {author} {\bibinfo {author} {\bibfnamefont {M.}~\bibnamefont
  {Schmidt}}, \bibinfo {author} {\bibfnamefont {S.}~\bibnamefont {Ke{\ss}ler}},
  \bibinfo {author} {\bibfnamefont {V.}~\bibnamefont {Peano}}, \bibinfo
  {author} {\bibfnamefont {O.}~\bibnamefont {Painter}}, \ and\ \bibinfo
  {author} {\bibfnamefont {F.}~\bibnamefont {Marquardt}},\ }\href
  {\doibase 10.1364/OPTICA.2.000635} {\bibfield  {journal} {\bibinfo
  {journal} {Optica}\ }\textbf {\bibinfo {volume} {2}},\ \bibinfo
  {eid} {635--641} (\bibinfo {year} {2015})}\BibitemShut {NoStop}%
  \bibitem [{\citenamefont {Xuereb}\ \emph {et~al.}(2015)\citenamefont {Xuereb},
  \citenamefont {Imparato},\ and\ \citenamefont
  {Dantan}}]{2015_Xuereb_HeatTransport_Array}%
  \BibitemOpen
  \bibfield  {author} {\bibinfo {author} {\bibfnamefont {A.}~\bibnamefont
  {Xuereb}}, \bibinfo {author} {\bibfnamefont {A.}~\bibnamefont {Imparato}}, \
  and\ \bibinfo {author} {\bibfnamefont {A.}~\bibnamefont {Dantan}},\ }\href
  {\doibase 10.1088/1367-2630/17/5/055013} {\bibfield  {journal} {\bibinfo
  {journal} {New J. Phys.}\ }\textbf {\bibinfo {volume} {17}},\ \bibinfo
  {pages} {055013} (\bibinfo {year} {2015})}\BibitemShut {NoStop}%
\bibitem [{\citenamefont {Peano}\ \emph {et~al.}(2014)\citenamefont {Peano},
  \citenamefont {Brendel}, \citenamefont {Schmidt},\ and\ \citenamefont
  {Marquardt}}]{2014_Peano_TopologicalPhases}%
  \BibitemOpen
 \bibfield  {author} {\bibinfo {author} {\bibfnamefont {V.}~\bibnamefont
  {Peano}}, \bibinfo {author} {\bibfnamefont {C.}~\bibnamefont {Brendel}},
  \bibinfo {author} {\bibfnamefont {M.}~\bibnamefont {Schmidt}}, \ and\
  \bibinfo {author} {\bibfnamefont {F.}~\bibnamefont {Marquardt}},\ }\href
  {\doibase 10.1103/PhysRevX.5.031011} {\bibfield  {journal} {\bibinfo
  {journal} {Phys. Rev. X}\ }\textbf {\bibinfo {volume} {5}},\ \bibinfo
  {eid} {031011} (\bibinfo {year} {2015})}\BibitemShut {NoStop}%
  \bibitem [{\citenamefont {Bhattacharya}\ and\ \citenamefont
  {Meystre}(2008)}]{2008_Bhattacharya_OM_multiMembrane}%
  \BibitemOpen
  \bibfield  {author} {\bibinfo {author} {\bibfnamefont {M.}~\bibnamefont
  {Bhattacharya}}\ and\ \bibinfo {author} {\bibfnamefont {P.}~\bibnamefont
  {Meystre}},\ }\href {\doibase 10.1103/PhysRevA.78.041801} {\bibfield
  {journal} {\bibinfo  {journal} {Phys. Rev. A}\ }\textbf {\bibinfo {volume}
  {78}},\ \bibinfo {pages} {041801} (\bibinfo {year} {2008})}\BibitemShut
  {NoStop}%
\bibitem [{\citenamefont {Xuereb}\ \emph {et~al.}(2012)\citenamefont {Xuereb},
  \citenamefont {Genes},\ and\ \citenamefont
  {Dantan}}]{2012_Xuereb_MembraneOptomechanicalArray}%
  \BibitemOpen
  \bibfield  {author} {\bibinfo {author} {\bibfnamefont {A.}~\bibnamefont
  {Xuereb}}, \bibinfo {author} {\bibfnamefont {C.}~\bibnamefont {Genes}}, \
  and\ \bibinfo {author} {\bibfnamefont {A.}~\bibnamefont {Dantan}},\ }\href
  {\doibase 10.1103/PhysRevLett.109.223601} {\bibfield  {journal} {\bibinfo
  {journal} {Phys. Rev. Lett.}\ }\textbf {\bibinfo {volume} {109}},\ \bibinfo
  {pages} {223601} (\bibinfo {year} {2012})}\BibitemShut {NoStop}%
\bibitem [{\citenamefont {Xuereb}\ \emph {et~al.}(2013)\citenamefont {Xuereb},
  \citenamefont {Genes},\ and\ \citenamefont
  {Dantan}}]{2013_Xuereb_EnhancedOMCoupling_Array}%
  \BibitemOpen
  \bibfield  {author} {\bibinfo {author} {\bibfnamefont {A.}~\bibnamefont
  {Xuereb}}, \bibinfo {author} {\bibfnamefont {C.}~\bibnamefont {Genes}}, \
  and\ \bibinfo {author} {\bibfnamefont {A.}~\bibnamefont {Dantan}},\ }\href
  {\doibase 10.1103/PhysRevA.88.053803} {\bibfield  {journal} {\bibinfo
  {journal} {Phys. Rev. A}\ }\textbf {\bibinfo {volume} {88}},\ \bibinfo
  {pages} {053803} (\bibinfo {year} {2013})}\BibitemShut {NoStop}%
\bibitem [{\citenamefont {Xuereb}\ \emph {et~al.}(2014)\citenamefont {Xuereb},
  \citenamefont {Genes}, \citenamefont {Pupillo}, \citenamefont {Paternostro},\
  and\ \citenamefont {Dantan}}]{2014_Xuereb_OMArray_LongRangePhononDynamics}%
  \BibitemOpen
  \bibfield  {author} {\bibinfo {author} {\bibfnamefont {A.}~\bibnamefont
  {Xuereb}}, \bibinfo {author} {\bibfnamefont {C.}~\bibnamefont {Genes}},
  \bibinfo {author} {\bibfnamefont {G.}~\bibnamefont {Pupillo}}, \bibinfo
  {author} {\bibfnamefont {M.}~\bibnamefont {Paternostro}}, \ and\ \bibinfo
  {author} {\bibfnamefont {A.}~\bibnamefont {Dantan}},\ }\href {\doibase
  10.1103/PhysRevLett.112.133604} {\bibfield  {journal} {\bibinfo  {journal}
  {Phys. Rev. Lett.}\ }\textbf {\bibinfo {volume} {112}},\ \bibinfo {pages}
  {133604} (\bibinfo {year} {2014})}\BibitemShut {NoStop}%
\bibitem [{\citenamefont {Heinrich}\ \emph {et~al.}(2011)\citenamefont
  {Heinrich}, \citenamefont {Ludwig}, \citenamefont {Qian}, \citenamefont
  {Kubala},\ and\ \citenamefont
  {Marquardt}}]{2011_Heinrich_CollectiveDynamics}%
  \BibitemOpen
  \bibfield  {author} {\bibinfo {author} {\bibfnamefont {G.}~\bibnamefont
  {Heinrich}}, \bibinfo {author} {\bibfnamefont {M.}~\bibnamefont {Ludwig}},
  \bibinfo {author} {\bibfnamefont {J.}~\bibnamefont {Qian}}, \bibinfo {author}
  {\bibfnamefont {B.}~\bibnamefont {Kubala}}, \ and\ \bibinfo {author}
  {\bibfnamefont {F.}~\bibnamefont {Marquardt}},\ }\href {\doibase
  10.1103/PhysRevLett.107.043603} {\bibfield  {journal} {\bibinfo  {journal}
  {Phys. Rev. Lett.}\ }\textbf {\bibinfo {volume} {107}},\ \bibinfo {pages}
  {043603} (\bibinfo {year} {2011})}\BibitemShut {NoStop}%
\bibitem [{\citenamefont {Kurths}\ \emph {et~al.}(2001)\citenamefont {Kurths},
  \citenamefont {Pikovsky},\ and\ \citenamefont
  {Rosenblum}}]{2001_Kurths_Synchronization}%
  \BibitemOpen
  \bibfield  {author} {\bibinfo {author} {\bibfnamefont {J.}~\bibnamefont
  {Kurths}}, \bibinfo {author} {\bibfnamefont {A.}~\bibnamefont {Pikovsky}}, \
  and\ \bibinfo {author} {\bibfnamefont {M.}~\bibnamefont {Rosenblum}},\
  }\href@noop {} {\emph {\bibinfo {title} {Synchronization: A Universal Concept
  in Nonlinear Sciences}}}\ (\bibinfo  {publisher} {Cambridge Universitz
  Press},\ \bibinfo {year} {2001})\BibitemShut {NoStop}%
  \bibitem [{\citenamefont {Giorgi}\ \emph {et~al.}(2012)\citenamefont {Giorgi},
  \citenamefont {Galve}, \citenamefont {Manzano}, \citenamefont {Colet},\ and\
  \citenamefont {Zambrini}}]{2012_Giorgi_QCorrAndSync}%
  \BibitemOpen
  \bibfield  {author} {\bibinfo {author} {\bibfnamefont {G.~L.}\ \bibnamefont
  {Giorgi}}, \bibinfo {author} {\bibfnamefont {F.}~\bibnamefont {Galve}},
  \bibinfo {author} {\bibfnamefont {G.}~\bibnamefont {Manzano}}, \bibinfo
  {author} {\bibfnamefont {P.}~\bibnamefont {Colet}}, \ and\ \bibinfo {author}
  {\bibfnamefont {R.}~\bibnamefont {Zambrini}},\ }\href {\doibase
  10.1103/PhysRevA.85.052101} {\bibfield  {journal} {\bibinfo  {journal} {Phys.
  Rev. A}\ }\textbf {\bibinfo {volume} {85}},\ \bibinfo {pages} {052101}
  (\bibinfo {year} {2012})}\BibitemShut {NoStop}%
\bibitem [{\citenamefont {{H\"ohberger}}\ and\ \citenamefont
  {Karrai}(2004)}]{2004_KarraiConstanze_IEEE}%
  \BibitemOpen
  \bibfield  {author} {\bibinfo {author} {\bibfnamefont {C.}~\bibnamefont
  {{H\"ohberger}}}\ and\ \bibinfo {author} {\bibfnamefont {K.}~\bibnamefont
  {Karrai}},\ }\href@noop {} {\bibfield  {journal} {\bibinfo  {journal}
  {Nanotechnology 2004, Proceedings of the 4th IEEE conference on
  nanotechnology,}\ \bibinfo {pages} {419}} (\bibinfo {year}
  {2004})}\BibitemShut {NoStop}%
\bibitem [{\citenamefont {Kippenberg}\ \emph {et~al.}(2005)\citenamefont
  {Kippenberg}, \citenamefont {Rokhsari}, \citenamefont {Carmon}, \citenamefont
  {Scherer},\ and\ \citenamefont {Vahala}}]{2005_KippenbergVahala_TheoryPRL}%
  \BibitemOpen
  \bibfield  {author} {\bibinfo {author} {\bibfnamefont {T.~J.}\ \bibnamefont
  {Kippenberg}}, \bibinfo {author} {\bibfnamefont {H.}~\bibnamefont
  {Rokhsari}}, \bibinfo {author} {\bibfnamefont {T.}~\bibnamefont {Carmon}},
  \bibinfo {author} {\bibfnamefont {A.}~\bibnamefont {Scherer}}, \ and\
  \bibinfo {author} {\bibfnamefont {K.~J.}\ \bibnamefont {Vahala}},\ }\href
  {http://journals.aps.org/prl/abstract/10.1103/PhysRevLett.95.033901}
  {\bibfield  {journal} {\bibinfo  {journal} {Phys. Rev. Lett.}\ }\textbf
  {\bibinfo {volume} {95}},\ \bibinfo {pages} {033901} (\bibinfo {year}
  {2005})}\BibitemShut {NoStop}%
\bibitem [{\citenamefont {Marquardt}\ \emph {et~al.}(2006)\citenamefont
  {Marquardt}, \citenamefont {Harris},\ and\ \citenamefont
  {Girvin}}]{2006_FM_DynamicalMultistability}%
  \BibitemOpen
  \bibfield  {author} {\bibinfo {author} {\bibfnamefont {F.}~\bibnamefont
  {Marquardt}}, \bibinfo {author} {\bibfnamefont {J.~G.~E.}\ \bibnamefont
  {Harris}}, \ and\ \bibinfo {author} {\bibfnamefont {S.~M.}\ \bibnamefont
  {Girvin}},\ }\href {\doibase 10.1103/PhysRevLett.96.103901} {\bibfield
  {journal} {\bibinfo  {journal} {Phys. Rev. Lett.}\ }\textbf {\bibinfo
  {volume} {96}},\ \bibinfo {eid} {103901} (\bibinfo {year}
  {2006})}\BibitemShut {NoStop}%
\bibitem [{\citenamefont {Metzger}\ \emph {et~al.}(2008)\citenamefont
  {Metzger}, \citenamefont {Ludwig}, \citenamefont {Neuenhahn}, \citenamefont
  {Ortlieb}, \citenamefont {Favero}, \citenamefont {Karrai},\ and\
  \citenamefont {Marquardt}}]{2008_Metzger}%
  \BibitemOpen
  \bibfield  {author} {\bibinfo {author} {\bibfnamefont {C.}~\bibnamefont
  {Metzger}}, \bibinfo {author} {\bibfnamefont {M.}~\bibnamefont {Ludwig}},
  \bibinfo {author} {\bibfnamefont {C.}~\bibnamefont {Neuenhahn}}, \bibinfo
  {author} {\bibfnamefont {A.}~\bibnamefont {Ortlieb}}, \bibinfo {author}
  {\bibfnamefont {I.}~\bibnamefont {Favero}}, \bibinfo {author} {\bibfnamefont
  {K.}~\bibnamefont {Karrai}}, \ and\ \bibinfo {author} {\bibfnamefont
  {F.}~\bibnamefont {Marquardt}},\ }\href {\doibase
  10.1103/PhysRevLett.101.133903} {\bibfield  {journal} {\bibinfo  {journal}
  {Phys. Rev. Lett.}\ }\textbf {\bibinfo {volume} {101}},\ \bibinfo {eid}
  {133903} (\bibinfo {year} {2008})}\BibitemShut {NoStop}%
\bibitem [{\citenamefont {Ludwig}\ \emph {et~al.}(2008)\citenamefont {Ludwig},
  \citenamefont {Kubala},\ and\ \citenamefont
  {Marquardt}}]{2008_LudwigMarquardt_OptomechInstab}%
  \BibitemOpen
  \bibfield  {author} {\bibinfo {author} {\bibfnamefont {M.}~\bibnamefont
  {Ludwig}}, \bibinfo {author} {\bibfnamefont {B.}~\bibnamefont {Kubala}}, \
  and\ \bibinfo {author} {\bibfnamefont {F.}~\bibnamefont {Marquardt}},\ }\href
  {http://stacks.iop.org/1367-2630/10/095013} {\bibfield  {journal} {\bibinfo
  {journal} {New J. Phys.}\ }\textbf {\bibinfo {volume} {10}},\ \bibinfo
  {pages} {095013} (\bibinfo {year} {2008})}\BibitemShut {NoStop}%
\bibitem [{\citenamefont {Qian}\ \emph {et~al.}(2012)\citenamefont {Qian},
  \citenamefont {Clerk}, \citenamefont {Hammerer},\ and\ \citenamefont
  {Marquardt}}]{2012_QianFM_NonclassicalStates}%
  \BibitemOpen
  \bibfield  {author} {\bibinfo {author} {\bibfnamefont {J.}~\bibnamefont
  {Qian}}, \bibinfo {author} {\bibfnamefont {A.}~\bibnamefont {Clerk}},
  \bibinfo {author} {\bibfnamefont {K.}~\bibnamefont {Hammerer}}, \ and\
  \bibinfo {author} {\bibfnamefont {F.}~\bibnamefont {Marquardt}},\ }\href
  {http://journals.aps.org/prl/abstract/10.1103/PhysRevLett.109.253601}
  {\bibfield  {journal} {\bibinfo  {journal} {Phys. Rev. Lett.}\ }\textbf
  {\bibinfo {volume} {109}},\ \bibinfo {pages} {253601} (\bibinfo {year}
  {2012})}\BibitemShut {NoStop}%
\bibitem [{\citenamefont {L\"orch}\ \emph {et~al.}(2014)\citenamefont
  {L\"orch}, \citenamefont {Qian}, \citenamefont {Clerk}, \citenamefont
  {Marquardt},\ and\ \citenamefont {Hammerer}}]{2014_Loerch_QLimitCycles}%
  \BibitemOpen
  \bibfield  {author} {\bibinfo {author} {\bibfnamefont {N.}~\bibnamefont
  {L\"orch}}, \bibinfo {author} {\bibfnamefont {J.}~\bibnamefont {Qian}},
  \bibinfo {author} {\bibfnamefont {A.}~\bibnamefont {Clerk}}, \bibinfo
  {author} {\bibfnamefont {F.}~\bibnamefont {Marquardt}}, \ and\ \bibinfo
  {author} {\bibfnamefont {K.}~\bibnamefont {Hammerer}},\ }\href {\doibase
  10.1103/PhysRevX.4.011015} {\bibfield  {journal} {\bibinfo  {journal} {Phys.
  Rev. X}\ }\textbf {\bibinfo {volume} {4}},\ \bibinfo {pages} {011015}
  (\bibinfo {year} {2014})}\BibitemShut {NoStop}%
\bibitem [{\citenamefont {Holmes}\ \emph {et~al.}(2012)\citenamefont {Holmes},
  \citenamefont {Meaney},\ and\ \citenamefont
  {Milburn}}]{2012_Holmes_Synchronization}%
  \BibitemOpen
  \bibfield  {author} {\bibinfo {author} {\bibfnamefont {C.~A.}\ \bibnamefont
  {Holmes}}, \bibinfo {author} {\bibfnamefont {C.~P.}\ \bibnamefont {Meaney}},
  \ and\ \bibinfo {author} {\bibfnamefont {G.~J.}\ \bibnamefont {Milburn}},\
  }\href {\doibase 10.1103/PhysRevE.85.066203} {\bibfield  {journal} {\bibinfo
  {journal} {Phys. Rev. E}\ }\textbf {\bibinfo {volume} {85}},\ \bibinfo
  {pages} {066203} (\bibinfo {year} {2012})}\BibitemShut {NoStop}%
\bibitem [{\citenamefont {Lauter}\ \emph {et~al.}(2015)\citenamefont {Lauter},
  \citenamefont {Brendel}, \citenamefont {Habraken},\ and\ \citenamefont
  {Marquardt}}]{2014_Lauter_PhasePatterns}%
  \BibitemOpen
  \bibfield  {author} {\bibinfo {author} {\bibfnamefont {R.}~\bibnamefont
  {Lauter}}, \bibinfo {author} {\bibfnamefont {C.}~\bibnamefont {Brendel}},
  \bibinfo {author} {\bibfnamefont {S.~J.~M.}\ \bibnamefont {Habraken}}, \ and\
  \bibinfo {author} {\bibfnamefont {F.}~\bibnamefont {Marquardt}},\ }\href
  {\doibase 10.1103/PhysRevE.92.012902} {\bibfield  {journal} {\bibinfo
  {journal} {Phys. Rev. E}\ }\textbf {\bibinfo {volume} {92}},\ \bibinfo
  {pages} {012902} (\bibinfo {year} {2015})}\BibitemShut {NoStop}%
\bibitem [{\citenamefont {Ludwig}\ and\ \citenamefont
  {Marquardt}(2013)}]{2013_Ludwig}%
  \BibitemOpen
  \bibfield  {author} {\bibinfo {author} {\bibfnamefont {M.}~\bibnamefont
  {Ludwig}}\ and\ \bibinfo {author} {\bibfnamefont {F.}~\bibnamefont
  {Marquardt}},\ }\href {\doibase 10.1103/PhysRevLett.111.073603} {\bibfield
  {journal} {\bibinfo  {journal} {Phys. Rev. Lett.}\ }\textbf {\bibinfo
  {volume} {111}},\ \bibinfo {pages} {073602} (\bibinfo {year}
  {2013})}\BibitemShut {NoStop}%
\bibitem [{\citenamefont {Walter}\ \emph
  {et~al.}(2014{\natexlab{a}})\citenamefont {Walter}, \citenamefont
  {Nunnenkamp},\ and\ \citenamefont {Bruder}}]{2014_Walter_QuantumSync_1Vdp}%
  \BibitemOpen
  \bibfield  {author} {\bibinfo {author} {\bibfnamefont {S.}~\bibnamefont
  {Walter}}, \bibinfo {author} {\bibfnamefont {A.}~\bibnamefont {Nunnenkamp}},
  \ and\ \bibinfo {author} {\bibfnamefont {C.}~\bibnamefont {Bruder}},\ }\href
  {\doibase 10.1103/PhysRevLett.112.094102} {\bibfield  {journal} {\bibinfo
  {journal} {Phys. Rev. Lett.}\ }\textbf {\bibinfo {volume} {112}},\ \bibinfo
  {pages} {094102} (\bibinfo {year} {2014}{\natexlab{a}})}\BibitemShut
  {NoStop}%
\bibitem [{\citenamefont {Walter}\ \emph
  {et~al.}(2014{\natexlab{b}})\citenamefont {Walter}, \citenamefont
  {Nunnenkamp},\ and\ \citenamefont
  {Bruder}}]{2014_Walter_SynchronizationVdPosc}%
  \BibitemOpen
  \bibfield  {author} {\bibinfo {author} {\bibfnamefont {S.}~\bibnamefont
  {Walter}}, \bibinfo {author} {\bibfnamefont {A.}~\bibnamefont {Nunnenkamp}},
  \ and\ \bibinfo {author} {\bibfnamefont {C.}~\bibnamefont {Bruder}},\ }\href
  {http://onlinelibrary.wiley.com/doi/10.1002/andp.201400144/abstract}
  {\bibfield  {journal} {\bibinfo  {journal} {Annalen der Physik}\ }\textbf
  {\bibinfo {volume} {527}},\ \bibinfo {pages} {131} (\bibinfo {year}
  {2014}{\natexlab{b}})}\BibitemShut {NoStop}%
\bibitem [{\citenamefont {Hermoso~de Mendoza}\ \emph
  {et~al.}(2014)\citenamefont {Hermoso~de Mendoza}, \citenamefont {Pach\'on},
  \citenamefont {G\'omez-Garde\~nes},\ and\ \citenamefont
  {Zueco}}]{2014_HermosoDeMendoza_SyncSemiclassicalKuramoto}%
  \BibitemOpen
  \bibfield  {author} {\bibinfo {author} {\bibfnamefont {I.}~\bibnamefont
  {Hermoso~de Mendoza}}, \bibinfo {author} {\bibfnamefont {L.~A.}\ \bibnamefont
  {Pach\'on}}, \bibinfo {author} {\bibfnamefont {J.}~\bibnamefont
  {G\'omez-Garde\~nes}}, \ and\ \bibinfo {author} {\bibfnamefont
  {D.}~\bibnamefont {Zueco}},\ }\href {\doibase 10.1103/PhysRevE.90.052904}
  {\bibfield  {journal} {\bibinfo  {journal} {Phys. Rev. E}\ }\textbf {\bibinfo
  {volume} {90}},\ \bibinfo {pages} {052904} (\bibinfo {year}
  {2014})}\BibitemShut {NoStop}%
\bibitem [{\citenamefont {Lee}\ and\ \citenamefont
  {Sadeghpour}(2013)}]{2013_Lee_QuantumSync_VdP_Ions}%
  \BibitemOpen
  \bibfield  {author} {\bibinfo {author} {\bibfnamefont {T.~E.}\ \bibnamefont
  {Lee}}\ and\ \bibinfo {author} {\bibfnamefont {H.~R.}\ \bibnamefont
  {Sadeghpour}},\ }\href {\doibase 10.1103/PhysRevLett.111.234101} {\bibfield
  {journal} {\bibinfo  {journal} {Phys. Rev. Lett.}\ }\textbf {\bibinfo
  {volume} {111}},\ \bibinfo {pages} {234101} (\bibinfo {year}
  {2013})}\BibitemShut {NoStop}%
\bibitem [{\citenamefont {Lee}\ \emph {et~al.}(2014)\citenamefont {Lee},
  \citenamefont {Chan},\ and\ \citenamefont
  {Wang}}]{2014_Lee_EntanglementTongue_QSync}%
  \BibitemOpen
  \bibfield  {author} {\bibinfo {author} {\bibfnamefont {T.~E.}\ \bibnamefont
  {Lee}}, \bibinfo {author} {\bibfnamefont {C.-K.}\ \bibnamefont {Chan}}, \
  and\ \bibinfo {author} {\bibfnamefont {S.}~\bibnamefont {Wang}},\ }\href
  {\doibase 10.1103/PhysRevE.89.022913} {\bibfield  {journal} {\bibinfo
  {journal} {Phys. Rev. E}\ }\textbf {\bibinfo {volume} {89}},\ \bibinfo
  {pages} {022913} (\bibinfo {year} {2014})}\BibitemShut {NoStop}%
\bibitem [{\citenamefont {Hush}\ \emph {et~al.}(2015)\citenamefont {Hush},
  \citenamefont {Li}, \citenamefont {Genway}, \citenamefont {Lesanovsky},\ and\
  \citenamefont {Armour}}]{2014_Hush_SpinCorr_QSync}%
  \BibitemOpen
  \bibfield  {author} {\bibinfo {author} {\bibfnamefont {M.~R.}\ \bibnamefont
  {Hush}}, \bibinfo {author} {\bibfnamefont {W.}~\bibnamefont {Li}}, \bibinfo
  {author} {\bibfnamefont {S.}~\bibnamefont {Genway}}, \bibinfo {author}
  {\bibfnamefont {I.}~\bibnamefont {Lesanovsky}}, \ and\ \bibinfo {author}
  {\bibfnamefont {A.~D.}\ \bibnamefont {Armour}},\ }\href {\doibase
  10.1103/PhysRevA.91.061401} {\bibfield  {journal} {\bibinfo  {journal} {Phys.
  Rev. A}\ }\textbf {\bibinfo {volume} {91}},\ \bibinfo {pages} {061401}
  (\bibinfo {year} {2015})}\BibitemShut {NoStop}%
\bibitem [{\citenamefont {Xu}\ \emph {et~al.}(2014)\citenamefont {Xu},
  \citenamefont {Tieri}, \citenamefont {Fine}, \citenamefont {Thompson},\ and\
  \citenamefont {Holland}}]{2014_Minghui_QuantumSync_Atoms}%
  \BibitemOpen
  \bibfield  {author} {\bibinfo {author} {\bibfnamefont {M.}~\bibnamefont
  {Xu}}, \bibinfo {author} {\bibfnamefont {D.~A.}\ \bibnamefont {Tieri}},
  \bibinfo {author} {\bibfnamefont {E.~C.}\ \bibnamefont {Fine}}, \bibinfo
  {author} {\bibfnamefont {J.~K.}\ \bibnamefont {Thompson}}, \ and\ \bibinfo
  {author} {\bibfnamefont {M.~J.}\ \bibnamefont {Holland}},\ }\href {\doibase
  10.1103/PhysRevLett.113.154101} {\bibfield  {journal} {\bibinfo  {journal}
  {Phys. Rev. Lett.}\ }\textbf {\bibinfo {volume} {113}},\ \bibinfo {pages}
  {154101} (\bibinfo {year} {2014})}\BibitemShut {NoStop}%
\bibitem [{\citenamefont {Zhirov}\ and\ \citenamefont
  {Shepelyansky}(2009)}]{2009_QSync_Qubits}%
  \BibitemOpen
  \bibfield  {author} {\bibinfo {author} {\bibfnamefont {O.~V.}\ \bibnamefont
  {Zhirov}}\ and\ \bibinfo {author} {\bibfnamefont {D.~L.}\ \bibnamefont
  {Shepelyansky}},\ }\href {\doibase 10.1103/PhysRevB.80.014519} {\bibfield
  {journal} {\bibinfo  {journal} {Phys. Rev. B}\ }\textbf {\bibinfo {volume}
  {80}},\ \bibinfo {pages} {014519} (\bibinfo {year} {2009})}\BibitemShut
  {NoStop}%
\bibitem [{\citenamefont {Savel'ev}\ \emph {et~al.}(2012)\citenamefont
  {Savel'ev}, \citenamefont {Washington}, \citenamefont {Zagoskin},\ and\
  \citenamefont {Everitt}}]{2012_Savelev_QSyncACdrives}%
  \BibitemOpen
  \bibfield  {author} {\bibinfo {author} {\bibfnamefont {S.~E.}\ \bibnamefont
  {Savel'ev}}, \bibinfo {author} {\bibfnamefont {Z.}~\bibnamefont
  {Washington}}, \bibinfo {author} {\bibfnamefont {A.~M.}\ \bibnamefont
  {Zagoskin}}, \ and\ \bibinfo {author} {\bibfnamefont {M.~J.}\ \bibnamefont
  {Everitt}},\ }\href {\doibase 10.1103/PhysRevA.86.065803} {\bibfield
  {journal} {\bibinfo  {journal} {Phys. Rev. A}\ }\textbf {\bibinfo {volume}
  {86}},\ \bibinfo {pages} {065803} (\bibinfo {year} {2012})}\BibitemShut
  {NoStop}%
\bibitem [{\citenamefont {Giorgi}\ \emph {et~al.}(2013)\citenamefont {Giorgi},
  \citenamefont {Plastina}, \citenamefont {Francica},\ and\ \citenamefont
  {Zambrini}}]{2013_Giorgi_QSync_Spins}%
  \BibitemOpen
  \bibfield  {author} {\bibinfo {author} {\bibfnamefont {G.~L.}\ \bibnamefont
  {Giorgi}}, \bibinfo {author} {\bibfnamefont {F.}~\bibnamefont {Plastina}},
  \bibinfo {author} {\bibfnamefont {G.}~\bibnamefont {Francica}}, \ and\
  \bibinfo {author} {\bibfnamefont {R.}~\bibnamefont {Zambrini}},\ }\href
  {\doibase 10.1103/PhysRevA.88.042115} {\bibfield  {journal} {\bibinfo
  {journal} {Phys. Rev. A}\ }\textbf {\bibinfo {volume} {88}},\ \bibinfo
  {pages} {042115} (\bibinfo {year} {2013})}\BibitemShut {NoStop}%
\bibitem [{\citenamefont {Hriscu}\ and\ \citenamefont
  {Nazarov}(2013)}]{2013_Hriscu_QSync_SuperconductingDevice}%
  \BibitemOpen
  \bibfield  {author} {\bibinfo {author} {\bibfnamefont {A.~M.}\ \bibnamefont
  {Hriscu}}\ and\ \bibinfo {author} {\bibfnamefont {Y.~V.}\ \bibnamefont
  {Nazarov}},\ }\href {\doibase 10.1103/PhysRevLett.110.097002} {\bibfield
  {journal} {\bibinfo  {journal} {Phys. Rev. Lett.}\ }\textbf {\bibinfo
  {volume} {110}},\ \bibinfo {pages} {097002} (\bibinfo {year}
  {2013})}\BibitemShut {NoStop}%
 \bibitem [{\citenamefont {Lee}\ \emph {et~al.}(2013)\citenamefont
  {Lee}, \ and\ \citenamefont
  {Vallade}}]{2013_Lee_CavSync_QCL}%
  \BibitemOpen
  \bibfield  {author} {\bibinfo {author} {\bibfnamefont {T.~E.}~\bibnamefont
  {Lee}},\ and\
  \bibinfo {author} {\bibfnamefont {M.~C.}\ \bibnamefont {Cross}},\ }\href
  {\doibase 10.1103/PhysRevA.88.013834} {\bibfield  {journal} {\bibinfo
  {journal} {Phys. Rev. A}\ }\textbf {\bibinfo {volume} {88}},\ \bibinfo
  {eid} {013834} (\bibinfo {year} {2013})}\BibitemShut {NoStop}%
\bibitem [{\citenamefont {Mari}\ \emph {et~al.}(2013)\citenamefont {Mari},
  \citenamefont {Farace}, \citenamefont {Didier}, \citenamefont {Giovannetti},\
  and\ \citenamefont {Fazio}}]{2013_Mari_QuantumSync}%
  \BibitemOpen
  \bibfield  {author} {\bibinfo {author} {\bibfnamefont {A.}~\bibnamefont
  {Mari}}, \bibinfo {author} {\bibfnamefont {A.}~\bibnamefont {Farace}},
  \bibinfo {author} {\bibfnamefont {N.}~\bibnamefont {Didier}}, \bibinfo
  {author} {\bibfnamefont {V.}~\bibnamefont {Giovannetti}}, \ and\ \bibinfo
  {author} {\bibfnamefont {R.}~\bibnamefont {Fazio}},\ }\href {\doibase
  10.1103/PhysRevLett.111.103605} {\bibfield  {journal} {\bibinfo  {journal}
  {Phys. Rev. Lett.}\ }\textbf {\bibinfo {volume} {111}},\ \bibinfo {pages}
  {103605} (\bibinfo {year} {2013})}\BibitemShut {NoStop}%
\bibitem [{\citenamefont {Ameri}\ \emph {et~al.}(2015)\citenamefont {Ameri},
  \citenamefont {Eghbali-Arani}, \citenamefont {Mari}, \citenamefont {Farace},
  \citenamefont {Kheirandish}, \citenamefont {Giovannetti},\ and\ \citenamefont
  {Fazio}}]{2015_Ameri_MutualInformation_QuantumSync}%
  \BibitemOpen
  \bibfield  {author} {\bibinfo {author} {\bibfnamefont {V.}~\bibnamefont
  {Ameri}}, \bibinfo {author} {\bibfnamefont {M.}~\bibnamefont
  {Eghbali-Arani}}, \bibinfo {author} {\bibfnamefont {A.}~\bibnamefont {Mari}},
  \bibinfo {author} {\bibfnamefont {A.}~\bibnamefont {Farace}}, \bibinfo
  {author} {\bibfnamefont {F.}~\bibnamefont {Kheirandish}}, \bibinfo {author}
  {\bibfnamefont {V.}~\bibnamefont {Giovannetti}}, \ and\ \bibinfo {author}
  {\bibfnamefont {R.}~\bibnamefont {Fazio}},\ }\href {\doibase
  10.1103/PhysRevA.91.012301} {\bibfield  {journal} {\bibinfo  {journal} {Phys.
  Rev. A}\ }\textbf {\bibinfo {volume} {91}},\ \bibinfo {pages} {012301}
  (\bibinfo {year} {2015})}\BibitemShut {NoStop}%
\bibitem [{\citenamefont {Bagheri}\ \emph {et~al.}(2013)\citenamefont
  {Bagheri}, \citenamefont {Poot}, \citenamefont {Fan}, \citenamefont
  {Marquardt},\ and\ \citenamefont {Tang}}]{2013_BagheriTang_Synchronization}%
  \BibitemOpen
  \bibfield  {author} {\bibinfo {author} {\bibfnamefont {M.}~\bibnamefont
  {Bagheri}}, \bibinfo {author} {\bibfnamefont {M.}~\bibnamefont {Poot}},
  \bibinfo {author} {\bibfnamefont {L.}~\bibnamefont {Fan}}, \bibinfo {author}
  {\bibfnamefont {F.}~\bibnamefont {Marquardt}}, \ and\ \bibinfo {author}
  {\bibfnamefont {H.~X.}\ \bibnamefont {Tang}},\ }\href {\doibase
  10.1103/PhysRevLett.111.213902} {\bibfield  {journal} {\bibinfo  {journal}
  {Phys. Rev. Lett.}\ }\textbf {\bibinfo {volume} {111}},\ \bibinfo {pages}
  {213902} (\bibinfo {year} {2013})}\BibitemShut {NoStop}%
\bibitem [{\citenamefont {Matheny}\ \emph {et~al.}(2014)\citenamefont
  {Matheny}, \citenamefont {Grau}, \citenamefont {Villanueva}, \citenamefont
  {Karabalin}, \citenamefont {Cross},\ and\ \citenamefont
  {Roukes}}]{2014_Matheny_PhaseSyncExp}%
  \BibitemOpen
  \bibfield  {author} {\bibinfo {author} {\bibfnamefont {M.~H.}\ \bibnamefont
  {Matheny}}, \bibinfo {author} {\bibfnamefont {M.}~\bibnamefont {Grau}},
  \bibinfo {author} {\bibfnamefont {L.~G.}\ \bibnamefont {Villanueva}},
  \bibinfo {author} {\bibfnamefont {R.~B.}\ \bibnamefont {Karabalin}}, \bibinfo
  {author} {\bibfnamefont {M.~C.}\ \bibnamefont {Cross}}, \ and\ \bibinfo
  {author} {\bibfnamefont {M.~L.}\ \bibnamefont {Roukes}},\ }\href {\doibase
  10.1103/PhysRevLett.112.014101} {\bibfield  {journal} {\bibinfo  {journal}
  {Phys. Rev. Lett.}\ }\textbf {\bibinfo {volume} {112}},\ \bibinfo {pages}
  {014101} (\bibinfo {year} {2014})}\BibitemShut {NoStop}%
\bibitem [{\citenamefont {Zhang}\ \emph {et~al.}(2015)\citenamefont {Zhang},
  \citenamefont {Shah}, \citenamefont {Cardenas},\ and\ \citenamefont
  {Lipson}}]{2015_Lipson_ArraySync}%
  \BibitemOpen
 \bibfield  {author} {\bibinfo {author} {\bibfnamefont {M.}~\bibnamefont
  {Zhang}}, \bibinfo {author} {\bibfnamefont {S.}~\bibnamefont {Shah}},
  \bibinfo {author} {\bibfnamefont {J.}~\bibnamefont {Cardenas}}, \ and\
  \bibinfo {author} {\bibfnamefont {M.}~\bibnamefont {Lipson}},\ }\href
  {\doibase 10.1103/PhysRevLett.115.163902} {\bibfield  {journal} {\bibinfo
  {journal} {Phys. Rev. Lett.}\ }\textbf {\bibinfo {volume} {115}},\ \bibinfo
  {eid} {163902} (\bibinfo {year} {2015})}\BibitemShut {NoStop}%
  %%%
   \bibitem [{\citenamefont {Huygens}\ \emph {}(1673)\citenamefont
  {Huygens}}]{1673_Huygens}%
  \BibitemOpen
  \bibfield  {author} {\bibinfo {author} {\bibfnamefont {C.}~\bibnamefont {Huygens}},\
  }\href@noop {} {\emph {\bibinfo {title} {Horologium Oscillatorium}}}\ (\bibinfo  {publisher} {Paris: Apud F. Muguet},\ \bibinfo {year} {1673}); English translation:  (Ames: Iowa State University Press, 1986)\BibitemShut {NoStop}%  
  %%%%%
\bibitem [{\citenamefont {Buczek}\ \emph {et~al.}(1972)\citenamefont
  {Buczek}, \ and\ \citenamefont
  {Freiberg}}]{1972_Buczek_AppliedSync_Lasers}%
  \BibitemOpen
  \bibfield  {author} {\bibinfo {author} {\bibfnamefont {C.~J.}~\bibnamefont
  {Buczek}}, \ and\
  \bibinfo {author} {\bibfnamefont {R.~J.}\ \bibnamefont {Freiberg}},\ }\href
  {\doibase 10.1109/JQE.1972.1077265} {\bibfield  {journal} {\bibinfo
  {journal} {IEEE J. Quantum Electron.}\ }\textbf {\bibinfo {volume} {8}},\ \bibinfo
  {eid} {641} (\bibinfo {year} {1972})}\BibitemShut {NoStop}%
    %%%%%
\bibitem [{\citenamefont {Cuomo}\ \emph {et~al.}(1993)\citenamefont
  {Cuomo}, \ and\ \citenamefont
  {Oppenheim}}]{1993_Cuomo_SyncAppl_Comm}%
  \BibitemOpen
  \bibfield  {author} {\bibinfo {author} {\bibfnamefont {K.~M.}~\bibnamefont
  {Cuomo}}, \ and\
  \bibinfo {author} {\bibfnamefont {A.~V.}\ \bibnamefont {Oppenheim}},\ }\href
  {\doibase 10.1103/PhysRevLett.71.65} {\bibfield  {journal} {\bibinfo
  {journal} {Phys. Rev. Lett.}\ }\textbf {\bibinfo {volume} {71}},\ \bibinfo
  {eid} {65} (\bibinfo {year} {1993})}\BibitemShut {NoStop}%
  %%%%
  \bibitem [{\citenamefont {Rosenblum}\ \emph {et~al.}(2003)\citenamefont
  {Rosenblum}, \ and\ \citenamefont
  {Pikovsky}}]{2003_Rosenblum_SyncRev}%
  \BibitemOpen
  \bibfield  {author} {\bibinfo {author} {\bibfnamefont {M.}~\bibnamefont
  {Rosenblum}}, \ and\
  \bibinfo {author} {\bibfnamefont {A.}\ \bibnamefont {Pikovsky}},\ } {\bibfield  {journal} {\bibinfo
  {journal} {Contemp. Phys.}\ }\textbf {\bibinfo {volume} {44}},\ \bibinfo
  {eid} {401} (\bibinfo {year} {2003})}\BibitemShut {NoStop}%
  %%%%
  \bibitem [{\citenamefont {Dörfler}\ \emph {et~al.}(2014)\citenamefont
  {Dörfler}, \ and\ \citenamefont
  {Bullo}}]{2014_Doerfler_SyncRev}%
  \BibitemOpen
  \bibfield  {author} {\bibinfo {author} {\bibfnamefont {F.}~\bibnamefont
  {Dörfler}}, \ and\
  \bibinfo {author} {\bibfnamefont {F.}\ \bibnamefont {Bullo}},\ }\href
  {\doibase 10.1016/j.automatica.2014.04.012} {\bibfield  {journal} {\bibinfo
  {journal} {Automatica}\ }\textbf {\bibinfo {volume} {50}},\ \bibinfo
  {eid} {1539} (\bibinfo {year} {2014})}\BibitemShut {NoStop}%
  %%%
    \bibitem [{\citenamefont {Samoilov}\ \emph {et~al.}(2004)\citenamefont
  {Samoilov}, \citenamefont {Plyasunov}, \ and\ \citenamefont
  {Arkin}}]{2004_NoiseIndBistab_bio}%
  \BibitemOpen
  \bibfield  {author} {\bibinfo {author} {\bibfnamefont {M.}~\bibnamefont
  {Samoilov}}, \bibinfo {author} {\bibfnamefont {S.}\ \bibnamefont {Plyasunov}}, \ and\
  \bibinfo {author} {\bibfnamefont {A.~P.}\ \bibnamefont {Arkin}},\ }\href
  {\doibase     10.1073/pnas.0406841102} {\bibfield  {journal} {\bibinfo
  {journal} {PNAS}\ }\textbf {\bibinfo {volume} {102}},\ \bibinfo
  {eid} {2310} (\bibinfo {year} {2004})}\BibitemShut {NoStop}%
  %%%
  \bibitem [{\citenamefont {Biancalani}\ \emph {et~al.}(2014)\citenamefont
  {Biancalani}, \citenamefont {Dyson}, \ and\ \citenamefont
  {McKane}}]{2014_Biancalani_NoiseIndBistab_bio}%
  \BibitemOpen
  \bibfield  {author} {\bibinfo {author} {\bibfnamefont {T.}~\bibnamefont
  {Biancalani}}, \bibinfo {author} {\bibfnamefont {L.}\ \bibnamefont {Dyson}}, \ and\
  \bibinfo {author} {\bibfnamefont {A.~J.}\ \bibnamefont {McKane}},\ }\href
  {\doibase 10.1103/PhysRevLett.112.038101} {\bibfield  {journal} {\bibinfo
  {journal} {Phys. Rev. Lett.}\ }\textbf {\bibinfo {volume} {99}},\ \bibinfo
  {eid} {038101} (\bibinfo {year} {2014})}\BibitemShut {NoStop}%
    %%% 
       \bibitem [{\citenamefont {Houchmandzadeh}\ \emph {et~al.}(2015)\citenamefont
  {Houchmandzadeh}, \ and\ \citenamefont
  {Vallade}}]{2015_BistabNoise_exact}%
  \BibitemOpen
  \bibfield  {author} {\bibinfo {author} {\bibfnamefont {B.}~\bibnamefont
  {Houchmandzadeh}},\ and\
  \bibinfo {author} {\bibfnamefont {M.}\ \bibnamefont {Vallade}},\ }\href
  {\doibase 10.1103/PhysRevE.91.022115} {\bibfield  {journal} {\bibinfo
  {journal} {Phys. Rev. E}\ }\textbf {\bibinfo {volume} {91}},\ \bibinfo
  {eid} {022115} (\bibinfo {year} {2015})}\BibitemShut {NoStop}%   
  %%%
   %%%
     %%%
          \bibitem [{\citenamefont {Dykman}\ \emph {et~al.}(1988)\citenamefont
  {Dykman}, \ and\ \citenamefont
  {Smelyanskiy}}]{1988_Dykman_BistabQuantum}%
  \BibitemOpen
  \bibfield  {author} {\bibinfo {author} {\bibfnamefont {M.~I.}~\bibnamefont
  {Dykman}},\ and\
  \bibinfo {author} {\bibfnamefont {V.~N.}\ \bibnamefont {Smelyanskiy}},\ }\href
  {http://www.pa.msu.edu/~dykman/pub06/sov.physJETP_88.pdf} {\bibfield  {journal} {\bibinfo
  {journal} {Zh. Eksp. Teor. Fiz}\ }\textbf {\bibinfo {volume} {94}},\ \bibinfo
  {eid} {61} (\bibinfo {year} {1988})}\BibitemShut {NoStop}%
  %%%
    \bibitem [{\citenamefont {Peano}\ \emph {et~al.}(2006)\citenamefont
  {Peano}, \ and\ \citenamefont
  {Thorwart}}]{2006_Peano_Bistab_QuantumDuffing}%
  \BibitemOpen
  \bibfield  {author} {\bibinfo {author} {\bibfnamefont {V.}~\bibnamefont
  {Peano}},\ and\
  \bibinfo {author} {\bibfnamefont {M.}\ \bibnamefont {Thorwart}},\ }\href
  {http://www.sciencedirect.com/science/article/pii/S0301010405002636} {\bibfield  {journal} {\bibinfo
  {journal} {Chem. Phys.}\ }\textbf {\bibinfo {volume} {322}},\ \bibinfo
  {eid} {135} (\bibinfo {year} {2006})}\BibitemShut {NoStop}%
   %%%
       \bibitem [{\citenamefont {Ghobadi}\ \emph {et~al.}(2011)\citenamefont
  {Ghobadi},\citenamefont
  {Bahrampour}, \ and\ \citenamefont
  {Simon}}]{2011_Ghobadi_Qom_bistab}%
  \BibitemOpen
  \bibfield  {author} {\bibinfo {author} {\bibfnamefont {R.}~\bibnamefont
  {Ghobadi}}, \bibinfo {author} {\bibfnamefont {A.~R.}\ \bibnamefont {Bahrampour}},\ and\
  \bibinfo {author} {\bibfnamefont {C.}\ \bibnamefont {Simon}},\ }\href
  {\doibase 10.1103/PhysRevA.84.033846} {\bibfield  {journal} {\bibinfo
  {journal} {Phys. Rev. A}\ }\textbf {\bibinfo {volume} {84}},\ \bibinfo
  {eid} {033846} (\bibinfo {year} {2011})}\BibitemShut {NoStop}%
   %%%
      \bibitem [{\citenamefont {Fichthorn}\ \emph {et~al.}(1989)\citenamefont
  {Fichthorn}, \citenamefont {Gulari},  \ and\ \citenamefont
  {Ziff}}]{1989_Fichthorn_Bistab_MC}%
  \BibitemOpen
  \bibfield  {author} {\bibinfo {author} {\bibfnamefont {K.}~\bibnamefont
  {Fichthorn}}, \bibinfo {author} {\bibfnamefont {E.}\ \bibnamefont {Gulari}}, \ and\
  \bibinfo {author} {\bibfnamefont {R.}\ \bibnamefont {Ziff}},\ }\href
  {\doibase 10.1103/PhysRevLett.63.1527} {\bibfield  {journal} {\bibinfo
  {journal} {Phys. Rev. Lett.}\ }\textbf {\bibinfo {volume} {63}},\ \bibinfo
  {eid} {1527} (\bibinfo {year} {1989})}\BibitemShut {NoStop}%
    %%% 
          \bibitem [{\citenamefont {Kim}\ \emph {et~al.}(1996)\citenamefont
  {Kim}, \citenamefont {Park},  \ and\ \citenamefont
  {Ryu}}]{1996_Kim_Multistab_OscArr}%
  \BibitemOpen
  \bibfield  {author} {\bibinfo {author} {\bibfnamefont {S.}~\bibnamefont
  {Kim}}, \bibinfo {author} {\bibfnamefont {S.~H.}\ \bibnamefont {Park}}, \ and\
  \bibinfo {author} {\bibfnamefont {C.~S.}\ \bibnamefont {Ryu}},\ }\href
  {\doibase 10.1103/PhysRevLett.78.1616} {\bibfield  {journal} {\bibinfo
  {journal} {Phys. Rev. Lett.}\ }\textbf {\bibinfo {volume} {78}},\ \bibinfo
  {eid} {1616} (\bibinfo {year} {1996})}\BibitemShut {NoStop}%
  %%%
   \bibitem [{\citenamefont {Residori}\ \emph {et~al.}(2001)\citenamefont
  {Residori}, \citenamefont {Berthet}, \citenamefont {Roman}, \ and\ \citenamefont
  {Fauve}}]{2001_Residori_Bistab_SurfWaves}%
  \BibitemOpen
  \bibfield  {author} {\bibinfo {author} {\bibfnamefont {S.}~\bibnamefont
  {Residori}}, \bibinfo {author} {\bibfnamefont {R.}\ \bibnamefont {Berthet}},  \bibinfo {author} {\bibfnamefont {B.}\ \bibnamefont {Roman}},\ and\
  \bibinfo {author} {\bibfnamefont {S.}\ \bibnamefont {Fauve}},\ }\href
  {\doibase 10.1103/PhysRevLett.88.024502} {\bibfield  {journal} {\bibinfo
  {journal} {Phys. Rev. Lett.}\ }\textbf {\bibinfo {volume} {88}},\ \bibinfo
  {eid} {024502} (\bibinfo {year} {2014})}\BibitemShut {NoStop}%
    %%% 
\bibitem [{\citenamefont {Kuramoto}(1975)}]{1975_Kuramoto_original}%
  \BibitemOpen
  \bibfield  {author} {\bibinfo {author} {\bibfnamefont {Y.}~\bibnamefont
  {Kuramoto}},\ }in\ \href {\doibase 10.1007/BFb0013365} {\emph {\bibinfo
  {booktitle} {International Symposium on Mathematical Problems in Theoretical
  Physics}}},\ \bibinfo {series} {Lecture Notes in Physics}, Vol.~\bibinfo
  {volume} {39},\ \bibinfo {editor} {edited by\ \bibinfo {editor}
  {\bibfnamefont {H.}~\bibnamefont {Araki}}}\ (\bibinfo  {publisher} {Springer
  Berlin Heidelberg},\ \bibinfo {year} {1975})\ pp.\ \bibinfo {pages}
  {420--422}\BibitemShut {NoStop}%
\bibitem [{\citenamefont {Acebr\'on}\ \emph {et~al.}(2005)\citenamefont
  {Acebr\'on}, \citenamefont {Bonilla}, \citenamefont {P\'erez~Vicente},
  \citenamefont {Ritort},\ and\ \citenamefont
  {Spigler}}]{2005_Acebron_KuramotoReview}%
  \BibitemOpen
  \bibfield  {author} {\bibinfo {author} {\bibfnamefont {J.~A.}\ \bibnamefont
  {Acebr\'on}}, \bibinfo {author} {\bibfnamefont {L.~L.}\ \bibnamefont
  {Bonilla}}, \bibinfo {author} {\bibfnamefont {C.~J.}\ \bibnamefont
  {P\'erez~Vicente}}, \bibinfo {author} {\bibfnamefont {F.}~\bibnamefont
  {Ritort}}, \ and\ \bibinfo {author} {\bibfnamefont {R.}~\bibnamefont
  {Spigler}},\ }\href {\doibase 10.1103/RevModPhys.77.137} {\bibfield
  {journal} {\bibinfo  {journal} {Rev. Mod. Phys.}\ }\textbf {\bibinfo {volume}
  {77}},\ \bibinfo {pages} {137} (\bibinfo {year} {2005})}\BibitemShut
  {NoStop}%
\bibitem [{\citenamefont {M{\o}lmer}\ \emph {et~al.}(1993)\citenamefont
  {M{\o}lmer}, \citenamefont {Castin},\ and\ \citenamefont
  {Dalibard}}]{1993_Molmer_quantumJumps}%
  \BibitemOpen
  \bibfield  {author} {\bibinfo {author} {\bibfnamefont {K.}~\bibnamefont
  {M{\o}lmer}}, \bibinfo {author} {\bibfnamefont {Y.}~\bibnamefont {Castin}}, \   
  and\ \bibinfo {author} {\bibfnamefont {J.}~\bibnamefont {Dalibard}},\ }\href
  {http://dx.doi.org/10.1364/JOSAB.10.000524} {\bibfield  {journal} {\bibinfo
  {journal} {JOSA B}\ }\textbf {\bibinfo {volume} {10}},\ \bibinfo {pages} {524-538} (\bibinfo {year}
  {1993})}\BibitemShut {NoStop}%
\bibitem [{\citenamefont {Plenio}\ and\ \citenamefont
  {Knight}(1998)}]{1998_Plenio_quantumJumps}%
  \BibitemOpen
  \bibfield  {author} {\bibinfo {author} {\bibfnamefont {M.~B.}\ \bibnamefont
  {Plenio}}\ and\ \bibinfo {author} {\bibfnamefont {P.~L.}\ \bibnamefont
  {Knight}},\ }\href {\doibase 10.1103/RevModPhys.70.101} {\bibfield  {journal}
  {\bibinfo  {journal} {Rev. Mod. Phys.}\ }\textbf {\bibinfo {volume} {70}},\
  \bibinfo {pages} {101} (\bibinfo {year} {1998})}\BibitemShut {NoStop}%
\bibitem [{\citenamefont {Ludwig}\ \emph {et~al.}(2012)\citenamefont {Ludwig},
  \citenamefont {Safavi-Naeini}, \citenamefont {Painter},\ and\ \citenamefont
  {Marquardt}}]{2012_ML_EnhancedQuNonlinearities}%
  \BibitemOpen
  \bibfield  {author} {\bibinfo {author} {\bibfnamefont {M.}~\bibnamefont
  {Ludwig}}, \bibinfo {author} {\bibfnamefont {A.~H.}\ \bibnamefont
  {Safavi-Naeini}}, \bibinfo {author} {\bibfnamefont {O.}~\bibnamefont
  {Painter}}, \ and\ \bibinfo {author} {\bibfnamefont {F.}~\bibnamefont
  {Marquardt}},\ }\href {\doibase 10.1103/PhysRevLett.109.063601} {\bibfield
  {journal} {\bibinfo  {journal} {Phys. Rev. Lett.}\ }\textbf {\bibinfo
  {volume} {109}},\ \bibinfo {pages} {063601} (\bibinfo {year}
  {2012})}\BibitemShut {NoStop}%
\bibitem [{\citenamefont {Kronwald}\ \emph {et~al.}(2013)\citenamefont
  {Kronwald}, \citenamefont {Ludwig},\ and\ \citenamefont
  {Marquardt}}]{2013_Kronwald_PhotonStatistics}%
  \BibitemOpen
  \bibfield  {author} {\bibinfo {author} {\bibfnamefont {A.}~\bibnamefont
  {Kronwald}}, \bibinfo {author} {\bibfnamefont {M.}~\bibnamefont {Ludwig}}, \
  and\ \bibinfo {author} {\bibfnamefont {F.}~\bibnamefont {Marquardt}},\ }\href
  {\doibase 10.1103/PhysRevA.87.013847} {\bibfield  {journal} {\bibinfo
  {journal} {Phys. Rev. A}\ }\textbf {\bibinfo {volume} {87}},\ \bibinfo
  {pages} {013847} (\bibinfo {year} {2013})}\BibitemShut {NoStop}%
\bibitem [{\citenamefont {Akram}\ \emph {et~al.}(2013)\citenamefont {Akram},
  \citenamefont {Bowen},\ and\ \citenamefont
  {Milburn}}]{2013_Akram_QJapplication}%
  \BibitemOpen
  \bibfield  {author} {\bibinfo {author} {\bibfnamefont {U.}~\bibnamefont
  {Akram}}, \bibinfo {author} {\bibfnamefont {W.~P.}\ \bibnamefont {Bowen}}, \
  and\ \bibinfo {author} {\bibfnamefont {G.~J.}\ \bibnamefont {Milburn}},\
  }\href {\doibase 10.1088/1367-2630/15/9/093007} {\bibfield  {journal}
  {\bibinfo  {journal} {New J. Phys.}\ }\textbf {\bibinfo {volume} {15}},\
  \bibinfo {pages} {093007} (\bibinfo {year} {2013})}\BibitemShut {NoStop}%
\bibitem [{\citenamefont {Mirza}\ and\ \citenamefont {van
  Enk}(2014)}]{2014_Mirza_SinglePhotonSpectra_QJ}%
  \BibitemOpen
  \bibfield  {author} {\bibinfo {author} {\bibfnamefont {I.~M.}\ \bibnamefont
  {Mirza}}\ and\ \bibinfo {author} {\bibfnamefont {S.~J.}\ \bibnamefont {van
  Enk}},\ }\href {\doibase 10.1103/PhysRevA.90.043831} {\bibfield  {journal}
  {\bibinfo  {journal} {Phys. Rev. A}\ }\textbf {\bibinfo {volume} {90}},\
  \bibinfo {pages} {043831} (\bibinfo {year} {2014})}\BibitemShut {NoStop}%
\bibitem [{\citenamefont {{Cohen}}\ \emph {et~al.}(2015)\citenamefont
  {{Cohen}}, \citenamefont {{Meenehan}}, \citenamefont {{MacCabe}},
  \citenamefont {{Gr{\"o}blacher}}, \citenamefont {{Safavi-Naeini}},
  \citenamefont {{Marsili}}, \citenamefont {{Shaw}},\ and\ \citenamefont
  {{Painter}}}]{2015_Cohen_phononCounting_Painter}%
  \BibitemOpen
  \bibfield  {author} {\bibinfo {author} {\bibfnamefont {J.~D.}\ \bibnamefont
  {{Cohen}}}, \bibinfo {author} {\bibfnamefont {S.~M.}\ \bibnamefont
  {{Meenehan}}}, \bibinfo {author} {\bibfnamefont {G.~S.}\ \bibnamefont
  {{MacCabe}}}, \bibinfo {author} {\bibfnamefont {S.}~\bibnamefont
  {{Gr{\"o}blacher}}}, \bibinfo {author} {\bibfnamefont {A.~H.}\ \bibnamefont
  {{Safavi-Naeini}}}, \bibinfo {author} {\bibfnamefont {F.}~\bibnamefont
  {{Marsili}}}, \bibinfo {author} {\bibfnamefont {M.~D.}\ \bibnamefont
  {{Shaw}}}, \ and\ \bibinfo {author} {\bibfnamefont {O.}~\bibnamefont
  {{Painter}}},\ }\href
  {http://www.nature.com/nature/journal/v520/n7548/full/nature14349.html}
  {\bibfield  {journal} {\bibinfo  {journal} {Nature}\ }\textbf {\bibinfo
  {volume} {520}},\ \bibinfo {pages} {522} (\bibinfo {year}
  {2015})}\BibitemShut {NoStop}%
\bibitem [{\citenamefont {Ueda}\ \emph {et~al.}(1990)\citenamefont {Ueda},
  \citenamefont {Imoto},\ and\ \citenamefont
  {Ogawa}}]{1990_Ueda_QJ_nonHermititanEvol}%
  \BibitemOpen
  \bibfield  {author} {\bibinfo {author} {\bibfnamefont {M.}~\bibnamefont
  {Ueda}}, \bibinfo {author} {\bibfnamefont {N.}~\bibnamefont {Imoto}}, \ and\
  \bibinfo {author} {\bibfnamefont {T.}~\bibnamefont {Ogawa}},\ }\href
  {\doibase 10.1103/PhysRevA.41.3891} {\bibfield  {journal} {\bibinfo
  {journal} {Phys. Rev. A}\ }\textbf {\bibinfo {volume} {41}},\ \bibinfo
  {pages} {3891} (\bibinfo {year} {1990})}\BibitemShut {NoStop}%
\bibitem [{\citenamefont {Chan}\ \emph {et~al.}(2012)\citenamefont {Chan},
  \citenamefont {Safavi-Naeini}, \citenamefont {Hill}, \citenamefont
  {Meenehan},\ and\ \citenamefont
  {Painter}}]{2012_Chan_OptimizedOptomechanicalCrystalCavity}%
  \BibitemOpen
  \bibfield  {author} {\bibinfo {author} {\bibfnamefont {J.}~\bibnamefont
  {Chan}}, \bibinfo {author} {\bibfnamefont {A.~H.}\ \bibnamefont
  {Safavi-Naeini}}, \bibinfo {author} {\bibfnamefont {J.~T.}\ \bibnamefont
  {Hill}}, \bibinfo {author} {\bibfnamefont {S.}~\bibnamefont {Meenehan}}, \
  and\ \bibinfo {author} {\bibfnamefont {O.}~\bibnamefont {Painter}},\ }\href
  {\doibase 10.1063/1.4747726} {\bibfield  {journal} {\bibinfo  {journal}
  {Appl. Phys. Lett.}\ }\textbf {\bibinfo {volume} {101}},\ \bibinfo {eid}
  {081115} (\bibinfo {year} {2012})}\BibitemShut {NoStop}%
\bibitem [{\citenamefont {Teufel}\ \emph {et~al.}(2011)\citenamefont {Teufel},
  \citenamefont {Donner}, \citenamefont {Li}, \citenamefont {Harlow},
  \citenamefont {Allman}, \citenamefont {Cicak}, \citenamefont {Sirois},
  \citenamefont {Whittaker}, \citenamefont {Lehnert},\ and\ \citenamefont
  {Simmonds}}]{2011_Teufel_SidebandCooling_Nature}%
  \BibitemOpen
  \bibfield  {author} {\bibinfo {author} {\bibfnamefont {J.~D.}\ \bibnamefont
  {Teufel}}, \bibinfo {author} {\bibfnamefont {T.}~\bibnamefont {Donner}},
  \bibinfo {author} {\bibfnamefont {D.}~\bibnamefont {Li}}, \bibinfo {author}
  {\bibfnamefont {J.~W.}\ \bibnamefont {Harlow}}, \bibinfo {author}
  {\bibfnamefont {M.~S.}\ \bibnamefont {Allman}}, \bibinfo {author}
  {\bibfnamefont {K.}~\bibnamefont {Cicak}}, \bibinfo {author} {\bibfnamefont
  {A.~J.}\ \bibnamefont {Sirois}}, \bibinfo {author} {\bibfnamefont {J.~D.}\
  \bibnamefont {Whittaker}}, \bibinfo {author} {\bibfnamefont {K.~W.}\
  \bibnamefont {Lehnert}}, \ and\ \bibinfo {author} {\bibfnamefont {R.~W.}\
  \bibnamefont {Simmonds}},\ }\href {http://dx.doi.org/10.1038/nature10261}
  {\bibfield  {journal} {\bibinfo  {journal} {Nature}\ }\textbf {\bibinfo
  {volume} {475}},\ \bibinfo {pages} {359} (\bibinfo {year}
  {2011})}\BibitemShut {NoStop}%
\bibitem [{\citenamefont {Murch}\ \emph {et~al.}(2008)\citenamefont {Murch},
  \citenamefont {Moore}, \citenamefont {Gupta},\ and\ \citenamefont
  {Stamper-Kurn}}]{2008_Murch_QuMeasurementBackaction}%
  \BibitemOpen
  \bibfield  {author} {\bibinfo {author} {\bibfnamefont {K.~W.}\ \bibnamefont
  {Murch}}, \bibinfo {author} {\bibfnamefont {K.~L.}\ \bibnamefont {Moore}},
  \bibinfo {author} {\bibfnamefont {S.}~\bibnamefont {Gupta}}, \ and\ \bibinfo
  {author} {\bibfnamefont {D.~M.}\ \bibnamefont {Stamper-Kurn}},\ }\href
  {http://dx.doi.org/10.1038/nphys965} {\bibfield  {journal} {\bibinfo
  {journal} {Nat. Phys.}\ }\textbf {\bibinfo {volume} {4}},\ \bibinfo {pages}
  {561} (\bibinfo {year} {2008})}\BibitemShut {NoStop}%
\bibitem [{\citenamefont {Walls}\ and\ \citenamefont
  {Milburn}(2008)}]{2008_WallsMilburn_QuantumOptics}%
  \BibitemOpen
  \bibfield  {author} {\bibinfo {author} {\bibfnamefont {D.}~\bibnamefont
  {Walls}}\ and\ \bibinfo {author} {\bibfnamefont {G.~J.}\ \bibnamefont
  {Milburn}},\ }\href@noop {} {\emph {\bibinfo {title} {Quantum Optics}}}\
  (\bibinfo  {publisher} {Springer-Verlag Berlin Heidelberg},\ \bibinfo {year}
  {2008})\BibitemShut {NoStop}%
\bibitem [{\citenamefont {Marquardt}\ \emph {et~al.}(2007)\citenamefont
  {Marquardt}, \citenamefont {Chen}, \citenamefont {Clerk},\ and\ \citenamefont
  {Girvin}}]{2007_FM_SidebandCooling}%
  \BibitemOpen
  \bibfield  {author} {\bibinfo {author} {\bibfnamefont {F.}~\bibnamefont
  {Marquardt}}, \bibinfo {author} {\bibfnamefont {J.~P.}\ \bibnamefont {Chen}},
  \bibinfo {author} {\bibfnamefont {A.~A.}\ \bibnamefont {Clerk}}, \ and\
  \bibinfo {author} {\bibfnamefont {S.~M.}\ \bibnamefont {Girvin}},\ }\href
  {\doibase 10.1103/PhysRevLett.99.093902} {\bibfield  {journal} {\bibinfo
  {journal} {Phys. Rev. Lett.}\ }\textbf {\bibinfo {volume} {99}},\ \bibinfo
  {eid} {093902} (\bibinfo {year} {2007})}\BibitemShut {NoStop}%
  %%%%%
  %%%%%
\end{thebibliography}

\end{document}